\documentclass[a4paper,10pt]{article}
\DeclareMathSizes{8}{8}{8}{8}
\usepackage[utf8x]{inputenc}
\usepackage{graphicx}
\usepackage{subfigure}
\usepackage{epsfig}
\usepackage{tabularx}
\usepackage{amsmath}
\usepackage{amssymb}
\usepackage[all]{xy} 
\usepackage{fancyhdr} 
\usepackage[numbers]{natbib} 
\newcommand{\beq}{\begin{equation}}
\newcommand{\eeq}{\end{equation}}
\newcommand{\beqa}{\begin{eqnarray}}
\newcommand{\eeqa}{\end{eqnarray}}        

\title{Modeling the coupled return-spread high frequency 
dynamics of large tick assets}
\author{Gianbiagio Curato \\
Scuola Normale Superiore di Pisa, Italy \\
\\
Fabrizio Lillo\\
Scuola Normale Superiore di Pisa,\\
Dipartimento di Fisica e Chimica, Universit\`a di Palermo, Italy}

\begin{document}
\bibliographystyle{plainnat} 

\maketitle

\begin{abstract}
Large tick assets, i.e. assets where one tick movement is a 
significant fraction of the price and bid-ask spread is almost 
always equal to one tick, display a dynamics in which price 
changes and spread are strongly coupled.  We introduce 
a Markov-switching modeling approach for price change, where the latent Markov process is the transition between spreads.  We  then use a finite Markov mixture of logit regressions on past squared returns to describe the dependence of the probability of price changes. The model can thus be seen as a Double Chain Markov Model.
We show that the model describes the shape of return distribution at different time aggregations, volatility clustering, and the anomalous decrease of kurtosis of returns. We calibrate our models on Nasdaq  stocks and we show that this model reproduces 
remarkably well the statistical properties of real data.

\end{abstract}

\bigskip

\bigskip

\break

\section{Introduction}
\label{sec:Introduction and motivation}
In financial markets, the price of an order cannot assume arbitrary values but it can be placed on a grid of values fixed by the exchange. The tick size is the smallest interval between two prices, i.e. the grid step, and it is measured in the currency of the asset. It is institutionally mandated and sets a limit on how finely prices may be specified.  The grid  is evenly spaced for a given asset, and the tick size depends on the price.

In the recent years there has been a growing interest toward the role of tick size in determining the statistical properties of returns, spread, limit order book, etc. \citep{Onnela,Munnix,Gillemot,Eisler,LaSpada,Rosenbaum,Dayri,Dayri2,Bouchaudnew}. 
The absolute tick size is not the best indicator for understanding and describing the high frequency dynamics of prices. Consider, for example, two highly liquid NASDAQ stocks, namely Apple (AAPL) and Microsoft (MSFT). For both stocks the tick size is one cent. However, in the period we investigated in this paper (July and August 2009), the average price of AAPL was 157\$ while the average price of MSFT was 24\$. Thus a one cent price movement for AAPL corresponds to 0.6 bp, while for MSFT it is 4.2 bp.  Therefore we can expect that the high frequency dynamics of AAPL will be significantly different from the one of MSFT. 
Recent literature has introduced the notion of an effective tick size to account and quantify the different
behavior of returns and spread processes of assets for a given value of tick size. Qualitatively we say that an asset
has a large tick size when the price is averse to variations of the order of a single tick and when the bid-ask spread is 
almost always equal to one tick. Conversely an asset is small tick size when the price is only weakly averse to variations of the order of a single tick and the bid-ask spread can assume a wide range of values, e.g. from one to ten or more ticks \citep{Rosenbaum,Dayri2}.
Several papers in empirical and theoretical market microstructure have emphasized that large and small tick size assets belong to different ``classes" \citep{Wyart,Eisler,Mike}. Order book models designed for small tick assets do not describe correctly the dynamics of large tick assets \citep{Mike}. Moreover the ultra high frequency statistical regularities of prices and of the order book are quite different in the two classes.

In this paper we are interested in modeling the dynamics of large tick assets at ultra high frequency and taking expliciteply into account the discreteness of prices. More specifically, we introduce a class of models describing the coupled dynamics of returns and spread for large tick assets in transaction time\footnote{Hereafter we define the transaction time as an integer counter of events defined by the execution of a market order. Note that if a market order is executed against several limit orders, our clock advances only by one unit.}. In our models, returns are defined as mid-price changes\footnote{With a little abuse of language we use returns and mid-price changes interchangeably .} and are measured in units of half tick, which is the minimum amount the mid-price can change. 
Therefore, these models are defined in a discrete state space \citep{McKenzie,Liesenfeld} and the time evolution is described in discrete time.
Our purpose is to model price dynamics in order to reproduce statistical properties of mid-price dynamics at different time scales and stylized facts like volatility clustering. Notice that, rather than considering a non observable efficient price and describing the data as the effect of the round off error due to tick size, we directly model the observable quantities, such as spread and mid-price, by using a time series approach.

\begin{figure}
	\centering
\includegraphics[width=0.49\textwidth]{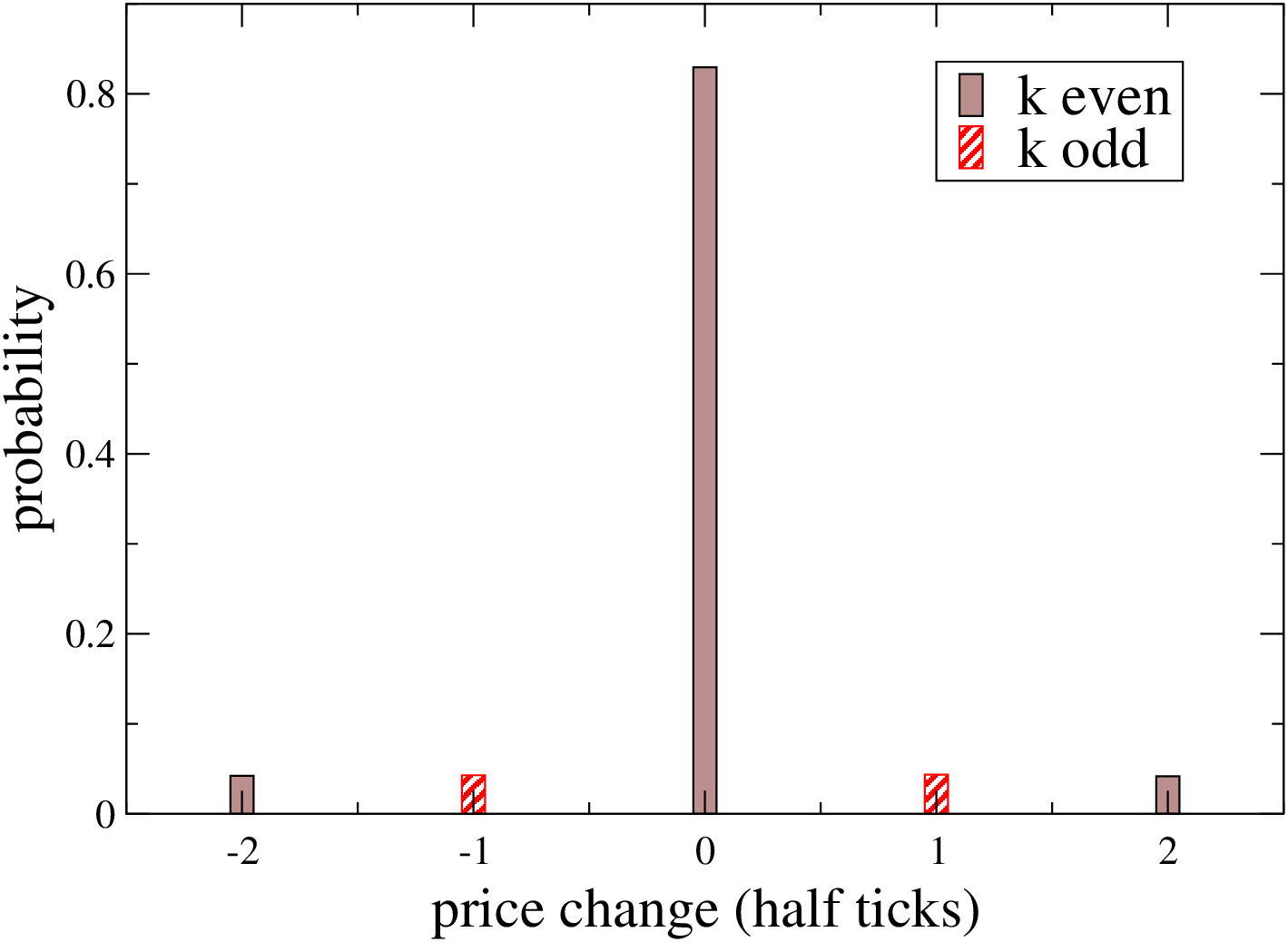}
\includegraphics[width=0.49\textwidth]{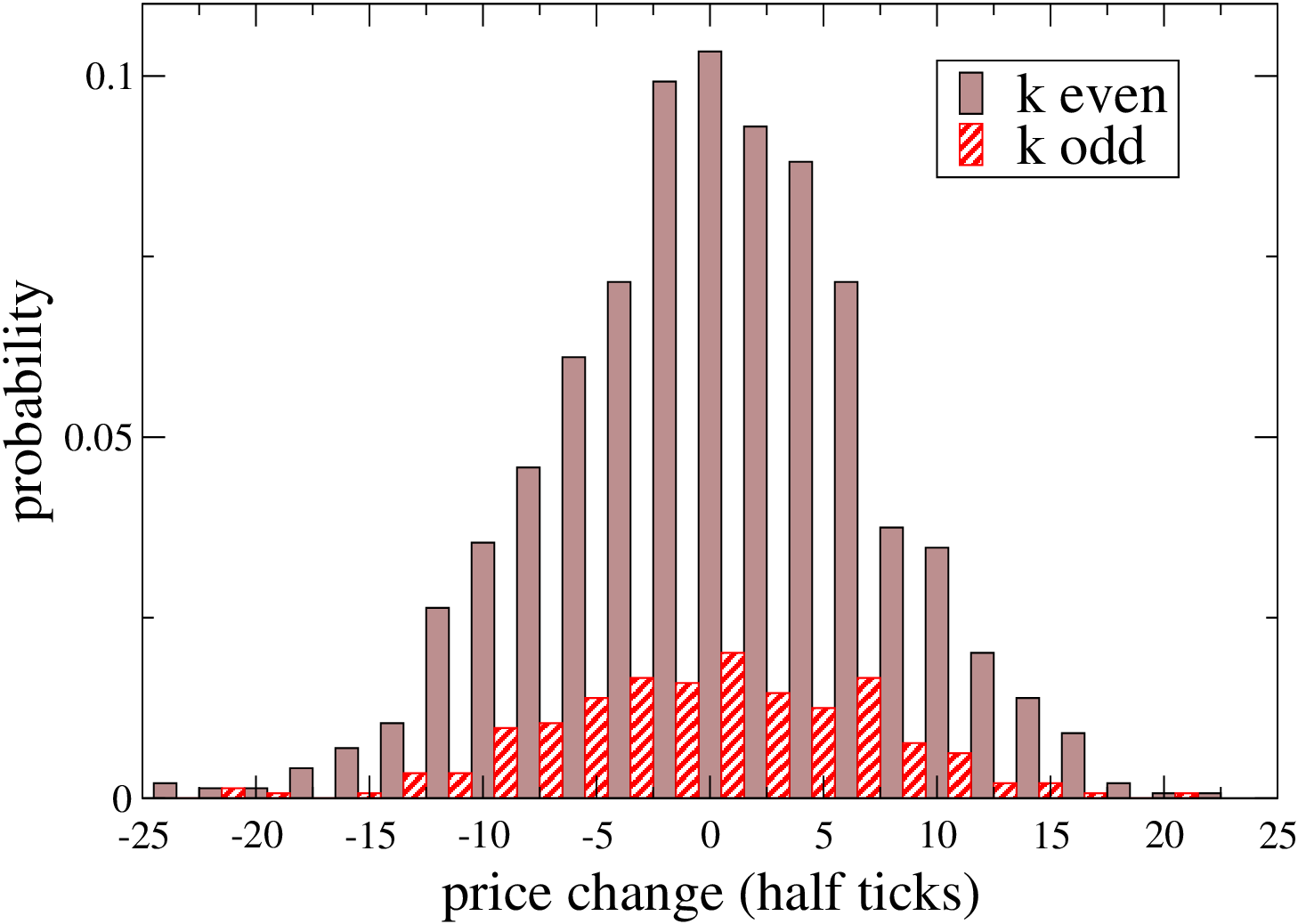}
	\caption{ The left panel
shows the tick by tick mid-price change distribution, $r\left(t,\Delta t=1\right)=p_{m}\left(t+1\right)-p_{m}\left(t\right)$,
while the right panel shows the mid-price change distribution aggregated at $128$ transactions,
$r\left(t,\Delta t=128\right)=p_{m}\left(t+128\right)-p_{m}\left(t\right)$. The investigated stock is Microsoft.}
	\label{fig:ret}
\end{figure}

The motivation of our work comes from two interesting empirical observations.
Let us consider first the unconditional distribution of mid price change at different time scales. In the left panel of Fig. \ref{fig:ret} we show the histogram of mid-price change of MSFT at the finest time scale, i.e. between two transactions. It is clear that most of the times the price does not change, while sometimes it changes by one or two half ticks. When we aggregate the returns on a longer time scale, for example $128$ transactions (see right panel of Fig. \ref{fig:ret}), a non trivial distribution emerges, namely a distribution where odd values of returns are systematically less populated than even values. It is important to notice that if we assume that returns of individual trades are independent and identically distributed\footnote{For example, if we randomize our sample of tick by tick mid-price changes}, we would never be able to reproduce an histogram like the one shown in the right panel of Fig. \ref{fig:ret}. In fact in this case the histogram would be, as expected, bell shaped.

\begin{figure}
	\centering
\includegraphics[width=0.55\textwidth]{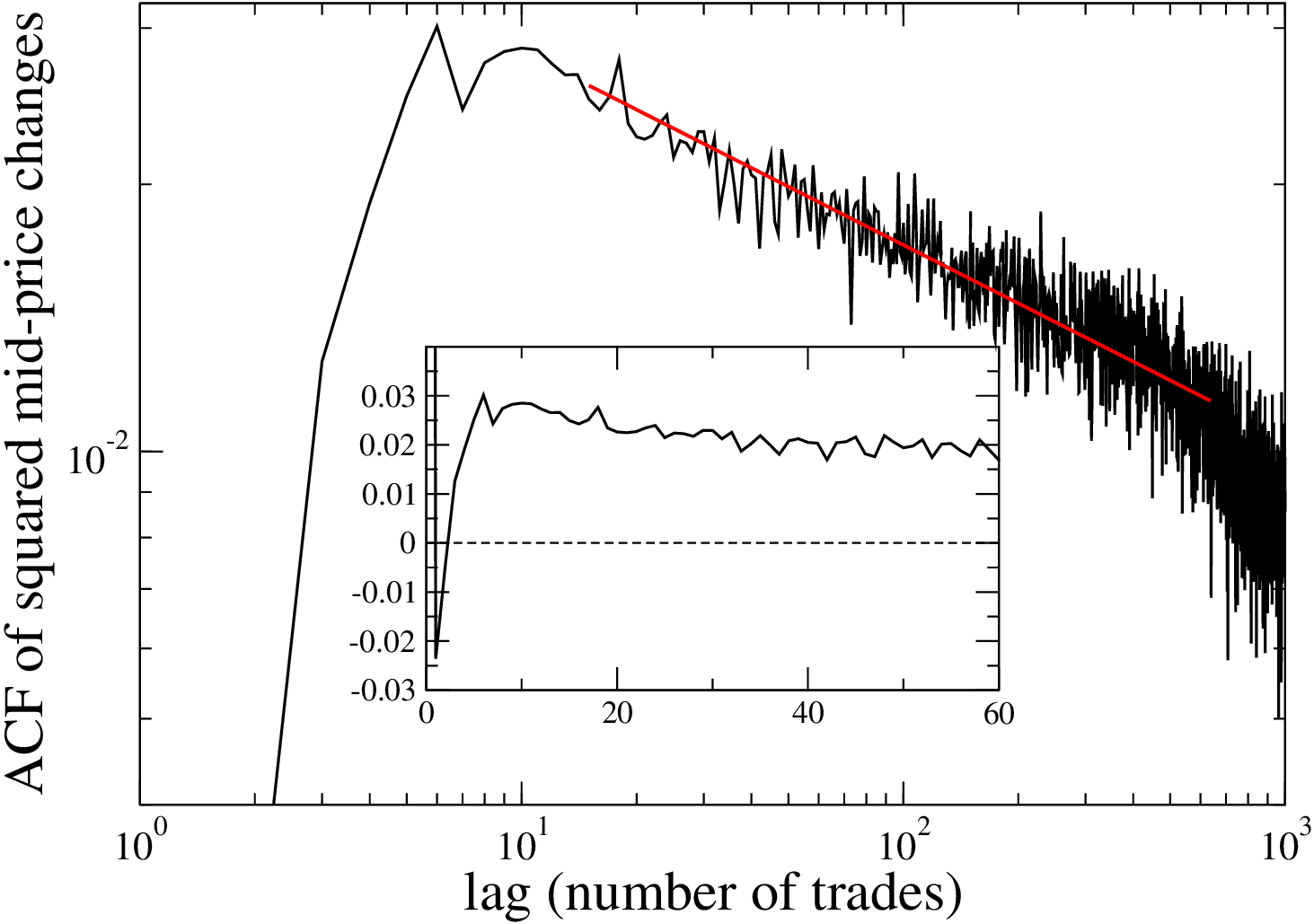}
	\caption{Sample autocorrelation function of tick by tick squared mid-price changes for Microsoft. The plot is in log-log scale and the red dashed line is a best fit of the autocorrelation function in the considered region. The estimated exponent is $\gamma=0.301$. The inset shows the behavior for small values of the lag.}
	\label{fig:volclust}
\end{figure}

The second observation concerns the properties of volatility of tick by tick returns. Figure \ref{fig:volclust} shows the autocorrelation function of squared returns of MSFT in transaction time. Square returns can be seen here as a simple proxy of volatility. First of all notice that the autocorrelation is negative for small lags. It then reaches a maximum around $10$ trades and then it decays very slowly to zero. We observe that between $10$ and more than $500$ trades, the decay of the autocorrelation function is well described by a power law function, $corr\left(r^{2}\left(t\right),r^{2}\left(t+\tau \right)\right)\sim \tau^{-\gamma}$, and the estimated exponent $\gamma\simeq 0.3$ is similar to the one observed at lower frequency and by sampling returns in real time rather than transaction time\footnote{It is worth noticing that in general the round-off error severely reduces the correlation properties of a stochastic process, even if the Hurst exponent of a long memory process is preserved \citep{LaSpada13}. Therefore the autocorrelation function shown in Fig. \ref{fig:volclust} is a strong underestimation of the tick by tick volatility clustering of the unobservable efficient price.}. We conclude therefore that very persisitent volatility clustering and possibly long range volatility is observed also at tick by tick level. 

The purpose of this paper is to develop a discrete time series model that is able to explain and reproduce simultaneously these two empirical observations, namely the change of the distribution of price changes at different time scales and the shape of the volatility autocorrelation.

As a modeling approach, we note that the observation of Fig. \ref{fig:ret} suggests that the return process can be characterized by different regimes which are defined by some variable, observable or not, in the order-book dynamics. The key intuition behind our modeling approach is that for large tick assets the dynamics of mid-price and of spread are intimately related and that the process of returns is conditioned to the spread process. 
The conditioning rule describes the connection between the stochastic motion of mid-price and spread on the grid. 

\begin{figure}
	\centering
		\includegraphics[width=0.7\textwidth]{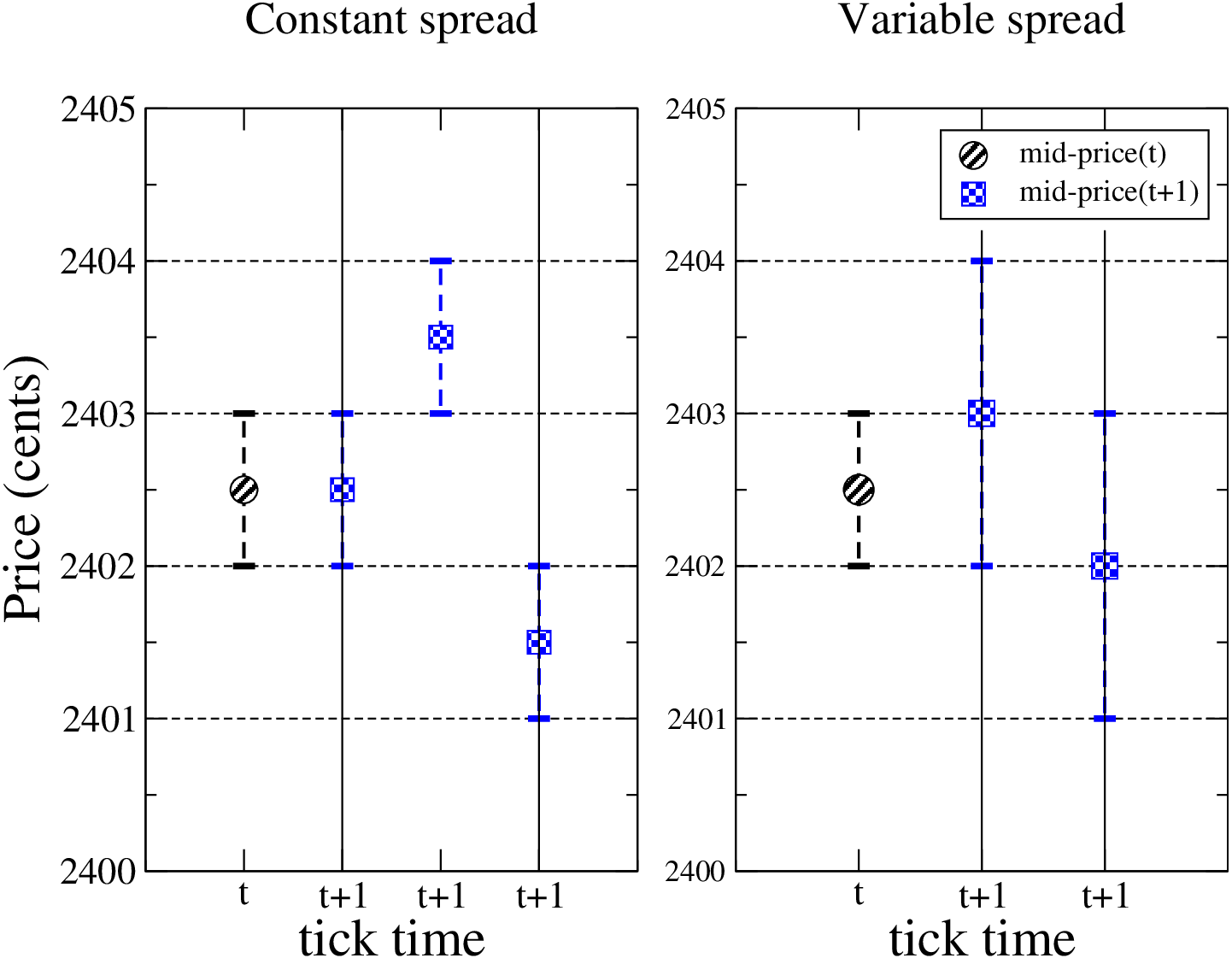}
	\caption{Coupling of spread and returns for large tick assets. On the left we show the three possible transitions when $s(t)=s(t+1)=1$. In this case the possible price changes are $r(t) \in \left(-2,0,2 \right)$ (measured
in 1/2 tick size). On the right we show the two possible transitions when $s(t)=1$ and $s(t+1)=2$. In this case the possible
values of price changes are $r(t) \in \left(-1,1 \right)$.}
	\label{fig:Transition}
\end{figure}

For large tick assets the spread typically assumes only few values. For example, for MSFT spread size is observed to be $1$ or $2$ ticks almost always.  The discreteness of mid-price dynamics can be connected to the spread dynamics if we observe that, when the spread is constant in time, returns can assume only even values. Instead when the spread changes, returns can display only odd values.  Figure \ref{fig:Transition} shows the mechanical relation between the two processes. The
dynamics of returns is thus linked to dynamics of spread transitions. This relation leads us to design models in which the return process depends on the transition between two subsequent spread states, distinguishing the case in which the spread remains constant and the case when it changes. From a methodological point of view we obtain this by defining a variable of state that describes the spread transition. We use a Hidden Markov, or Markov Switching, Model \citep{Rabiner,Hamilton} for returns, in which the spread transition is described by a Markov chain that defines different regimes for the return process. 


The Markov Switching approach is able to describe the change in shape of the distribution of price change (Fig. \ref{fig:ret}), but not the persistence of volatility. To this end, we propose a more sophisticated model by allowing the returns process to be an regressive process in which regressors are the past value of squared returns \citep{Wallach,Berchtold1,Berchtold2,Raftery}. 
We show how to calibrate the models on real data and we tested them on the large tick assets MSFT and CSCO, traded at NASDAQ market in the period July-August 2009. We show that the full model reproduces very well the empirical data.


The paper is organized as follows. In Section \ref{sec:Review of Markov Switching models in econometrics} we review the main applications of Markov-switching modeling in the econometrics field. 
In Section \ref{sec:Econometric model} we present our modeling approach. In Section \ref{sec:Data and stylized facts} we present our data for the MSFT stock and we describe the observed stylized facts of price dynamics. In Section \ref{sec:Simulation} we describe the calibration of the models on real data and we discuss how well the different models reproduce the stylized facts. Finally, in Section \ref{sec:conclusion} we draw some conclusions and we discuss future works. 

\section{Review of Markov switching models in econometrics}
\label{sec:Review of Markov Switching models in econometrics}

Markov switching models (MS models) have become increasingly popular in econometric studies of industrial production, interest rates, stock prices and unemployment
rates \citep{Hamilton,Schnatter}. They are also known as hidden Markov models (HMM) \citep{Rabiner,Ryden,Bulla}, used for example in speech recognition and DNA analysis. In these models the distribution that generates an observation depends on the states of an 
underlying and unobserved Markov process. They are flexible general purpose models for univariate and multivariate time series, especially for discrete-valued series, including categorical variables and series of counts \citep{Zucchini}.  
Markov switching models belong to a general class of mixture distributions \citep{Schnatter}. Econometricians' initial interest in this class of distributions was based on their ability to flexibly approximate general classes of density functions and generate a wider range of values
for the skewness and kurtosis than is obtainable by using a single distribution. Along these lines Granger and Orr \citep{Granger} and Clark \citep{Clark} considered time-independent mixtures of normal distributions as a means of modeling non-normally distributed data.
These models, however, did not capture the time dependence in the conditional variance found in many economic time series, as evidenced by the vast literature on ARCH models that started with Engle \citep{Hamilton}.   
By allowing the mixing probabilities to display time dependence, Markov switching models can be seen as a natural generalization of the original time-independent mixture of normals model. Timmermann \citep{Timmermann} has shown that the mixing property enables them to generate a wide range of 
coefficients of skewness, kurtosis and serial correlation even when based on a very small number of underlying states. 
Regime switches in economic time series can be parsimoniously represented by Markov switching models by letting the mean, variance, and possibly the dynamics of the series depend on the realization of a finite number of discrete states.

The basic MS model is:
\begin{equation}
\label{eq:4.1}
y\left(t\right)=\mu_{S\left(t\right)}+\sigma_{S\left(t\right)}\epsilon \left(t\right), 
\end{equation} 
where $S\left(t\right)=1,2,\cdots,k$ denotes the unobserved state indicator which follows an ergodic $k$-state Markov process and
$\epsilon \left(t\right)$ is a zero-mean random variable which is i.i.d. over time \citep{Engle}. Another relevant model is the Markov 
switching autoregressive model (MSAR($q$)) of order $q$ that allows for state-independent autoregressive dynamics:
\begin{equation}
\label{eq:4.2}
y\left(t\right)=\mu_{S\left(t\right)}+\sum_{j=1}^q \phi_{j}\left(y\left(t-j\right)-\mu_{S\left(t-j\right)}\right)+\sigma_{S\left(t\right)}\epsilon \left(t\right).
\end{equation}
It became popular in econometrics for analyzing economic time series such as the GDP data through the work of Hamilton \citep{Hamilton2}. In its most general form the MSAR model
allows that the autoregressive coefficients are also affected by $S\left(t\right)$ \citep{Timmermann}:
\begin{equation}
\label{eq:4.2}
y\left(t\right)=\mu_{S\left(t\right)}+\sum_{j=1}^q \phi_{j,S\left(t-j\right)}\left(y\left(t-j\right)-\mu_{S\left(t-j\right)}\right)+\sigma_{S\left(t\right)}\epsilon \left(t\right).
\end{equation}

There is a key difference with respect to ARCH models, which is another type of time-dependent mixture processes. While Markov switching models mix a finite number of states with different mean and
volatility parameters based on an exogenous state process, ARCH models mix distributions with volatility parameters drawn from an infinite set of states driven by lagged innovations to the series. 

We can make use of the above models when we want to model a continuous state random variable $y\left(t\right)$. In our case we want a model for a discrete variable, i.e. the observed integer price differences, in a
microstructure market environment. 
Therefore the models for continuous variables presented above cannot be used in our problem. We propose to model the coupled dynamics of spreads and price differences in the setting defined by the Double
Chain Markov Models (DCMM) \citep{Berchtold1,Berchtold2}. This is the natural extension of HMM models in order to allow the hidden Markov process to select one of a finite number of Markov chains
to drive the observed process at each time point. If a time series can be decomposed into a finite mixture of Markov chains, then the
DCMM can be applied to describe the switching process between these chains. In turn DCMM belongs to the family of Markov chains in random environments \citep{Cogburn,Cogburn2}.  

In discrete time, DCCM describes the joint dynamics of two random variables: $x\left(t\right)$, whose state at time $t$ is unknown for an observer external to the process, and $y\left(t\right)$, which is observable.
The model is described by the following elements:
\begin{itemize}
 \item A set of hidden states, $\mathcal{S}\left(x\right)=\left \{ 1,\cdots,N_{x}\right \}$.
 \item A set of possible outputs, $\mathcal{S}\left(y\right)=\left \{ 1,\cdots,N_{y}\right \}$.
 \item The probability distribution of the first hidden state, $\boldsymbol{\pi_{0}}=\left \{\pi_{0,1},\cdots,\pi_{0,N_{x}}\right \}$.
 \item A transition matrix between hidden states, $M=\left \{m_{ij}\right \},\ i,\ j \in \mathcal{S}\left( x\right)$.
 \item A set of transition matrices between successive outputs of $y\left(t\right)$ given a particular state of $x\left(t\right)$,
 $V_{x\left(t\right)=k,ij},\ i,\ j \in \mathcal{S} \left(y\right)$, $k \in \mathcal{S} \left(x\right)$. 
\end{itemize}
There are three different estimation problems: the estimation of the probability of a sequence of observations $y(0),\cdots,y(T)$ given a model;
the estimation of parameters $\boldsymbol{\pi_{0}},\ M,\ V_{k}$ given a sequence of observations; the estimation of the optimal sequence of hidden states
given a model and a sequence of outputs. 

Our data, i.e. limit order book data, instead allow us to see directly the process that defines the hidden Markov process, i.e. the spread process.
In this way we can estimate directly the matrices $M$ and $V_{k}$ by a simple maximum likelihood approach without using the Expectation Maximization (EM)
algorithm and the Viterbi algorithm, that are usually used when the hidden process is not observable \citep{Berchtold1,Berchtold2}. We use the stationary probability distribution for the process $x\left(t\right)$ 
as initial probability distribution $\boldsymbol{\pi_{0}}$ in order to perform our calculations and simulations.
We use the DCMM model as a mathematical framework for spread and price differences processes without treating spread process as an hidden process. 

Among the few financial applications of the DCMM model we mention Ref.s \citep{Fitzpatrick,Eisenkopf}. In the former paper, authors studied the credit rating dynamics of a portfolio of financial companies, where the unobserved hidden process is the state of the broader economy. In Eisenkopf \citep{Eisenkopf} instead the author considered a problem in which a 
credit rating process is influenced by unobserved hidden risk situations. To the best of our knowledge our paper is the first application of DCMM to the field of market microstructure and high frequency financial data. 

\section{Models for the coupled dynamics of spread and returns}
\label{sec:Econometric model}

In this section we present the models describing the process of
returns $r\left ( t,\Delta t \right )= p_{m}\left ( t +\Delta t\right
)-p_{m}\left ( t\right )$ at time scale $\Delta t$, where we define the 
mid-price as $p_{m}\left ( t  \right)=\left ( p_{ASK}\left ( t 
\right)+p_{BID}\left ( t  \right)  \right)/2$ and we choose to measure $r$ in units of half tick size. In our models, return process  follows different time series processes conditioned on the dynamics of transitions of the spread $s\left(t\right)=p_{ASK}\left ( t 
\right)-p_{BID}\left ( t  \right)$.
Hereafter we will use the notation $r\left (
t\right )=r\left ( t,\Delta t=1 \right )$.
The spread variable $s$ is measured in units of $1$ tick size, so we have
$r\left ( t,  \Delta t\right)\in \mathbb{Z}$ and $s\left ( t  \right)\in \mathbb{N}$. The
time variable $t \in \mathbb{N}$ is the transaction time.



\subsection{Markov-Switching models}
\label{sec:Markov-Switching models}

{\it Spread process.} It is well known that spread process is autocorrelated in time \citep{Plerou,Bouchaud,Hautsch,Ponzi}. 
We model the spread  $s\left ( t  \right)$ as a
stationary Markov(1) \citep{Pham} process\footnote{We have tried other specifications of the spread process, such as for example a long memory process, but this does not change significantly our results.}:
\begin{equation}
\label{eq:1.2}
P\left ( s\left ( t \right )=j|s\left ( t-1 \right )=i,s\left ( t-2 \right
)=k,\cdots \right )=P\left ( s\left ( t \right )=j|s\left ( t-1 \right )=i
\right )=p_{ij}, 
\end{equation}
where $i,j \in \mathbb{N}$ are spread values. As mentioned, we limit the set of spread values to $s \in \left
\{ 1,2 \right \}$, because we want to describe the case of large tick assets.
We also assume that the process $s\left(t\right)$ is not affected by the return process
$r\left(t\right)$. The spread process is described by the transition matrix:
$$
B=
\begin{pmatrix} 
p_{11} & p_{12} \\
p_{21} & p_{22} 
\end{pmatrix}
$$
where the normalization is given by $\sum_{j=1}^2 p_{ij}=1$. The vector of
stationary probabilities is the eigenvector $\boldsymbol{\pi}$ of $B'$
relative to eigenvalue $1$, which is
\begin{equation}
\label{eq:1.3}
B'\boldsymbol{\pi}=\boldsymbol{\pi},\quad  \quad \quad\boldsymbol{\pi}=
\begin{pmatrix}
\left ( 1-p_{22} \right )/\left ( 2-p_{11}-p_{22} \right ) \\
\left ( 1-p_{11} \right )/\left ( 2-p_{11}-p_{22} \right )
\end{pmatrix}
\end{equation}
where $B'$ denotes the transpose of the matrix $B$. This vector represents the unconditional probabilities of $s\left
( t  \right)$, so $\pi_{k}=P\left (s\left ( t \right )=k \right )$ with $k=1,2$.

Starting from the $s\left ( t  \right)$ process, it is useful to define a new stationary
Markov(1) process $x\left ( t  \right)$ that describes the stochastic
dynamics of 
transitions between states $s\left ( t  \right)$ and $s\left ( t+1  \right)$ as
\begin{eqnarray}
\label{eq:1.4}
 \ x\left ( t  \right)=1 \quad &{\mbox if}& \quad s\left ( t+1  \right)=1,s\left ( t 
\right)=1,     \nonumber \\
 \ x\left ( t  \right)=2 \quad &{\mbox if}& \quad s\left ( t+1  \right)=2,s\left ( t 
\right)=1,     \nonumber \\
 \ x\left ( t  \right)=3 \quad &{\mbox if}& \quad s\left ( t+1  \right)=1,s\left ( t 
\right)=2,     \nonumber \\
 \ x\left ( t  \right)=4 \quad &{\mbox if}& \quad s\left ( t+1  \right)=2,s\left ( t 
\right)=2.
\end{eqnarray}
This process is characterized by a new transition matrix
$$
M=
\begin{pmatrix} 
m_{11} & m_{12} & m_{13} & m_{14} \\
m_{21} & m_{22} & m_{23} & m_{24} \\
m_{31} & m_{32} & m_{33} & m_{34} \\
m_{41} & m_{42} & m_{43} & m_{44}
\end{pmatrix}
=
\begin{pmatrix} 
p_{11} & p_{12} & 0 & 0 \\
0 & 0 & p_{21} & p_{22} \\
p_{11} & p_{12} & 0 & 0 \\
0 & 0 & p_{21} & p_{22}  
\end{pmatrix}
$$
in which the stationary vector is given by
\begin{equation}
\label{eq:1.5}
M'\boldsymbol{\lambda}=\boldsymbol{\lambda},\quad \boldsymbol{\lambda}=
\begin{pmatrix}
\left ( p_{21}p_{11} \right )/\left ( 1-p_{11}+p_{21} \right ) \\
p_{21}\left ( 1-p_{11} \right )/\left ( 1-p_{11}+p_{21} \right ) \\
p_{21}\left ( 1-p_{11} \right )/\left ( 1-p_{11}+p_{21} \right ) \\
\left ( 1-p_{21} \right )\left ( 1-p_{11} \right )/\left ( 1-p_{11}+p_{21}
\right )
\end{pmatrix}.
\end{equation}

A limiting case is when the spread process $s\left ( t  \right)$
is described by a Bernoulli process. In this case we set  $P\left ( s\left
( t  \right)=1  \right)=p$.
Although $s\left ( t  \right)$ is an i.i.d. process, the spread transition process $x_{B}\left ( t  \right)$
is a Markov process defined by:
$$
M_{B}=
\begin{pmatrix} 
p & \left ( 1-p \right ) & 0 & 0 \\
0 & 0 & p & \left ( 1-p \right ) \\
p & \left ( 1-p \right ) & 0 & 0 \\
0 & 0 & p & \left ( 1-p \right )  
\end{pmatrix},
\quad \boldsymbol{\lambda}_{B}=
\begin{pmatrix}
p^2 \\
p\left ( 1-p \right ) \\
p\left ( 1-p \right ) \\
\left ( 1-p \right )^2
\end{pmatrix}.
$$

In the general case, the process $x\left ( t  \right)$ is defined by two parameters
$p_{11}, p_{21}$ (which are reduced to $p$ in Bernoulli case) that we can estimate from spread data.

\medskip

{\it Mid-price process.} We can now define a Markov-switching process for returns $r\left ( t  \right)$ which is conditioned to the process $x \left ( t  \right)$, i.e. to the spread transitions.
Returns are measured in half ticks and we limit the set of possible values to $r\left ( t  \right)\in \left \{ -2,-1,0,1,2
\right \}$, as observed in our sample. The discreteness of the price grid imposes the mechanical constraints 
\begin{eqnarray}
\label{eq:1.6}
 \ x\left ( t  \right)=1 \quad &\longrightarrow& r\left ( t  \right)\in \left
\{ -2,0,2 \right \},     \nonumber \\
 \ x\left ( t  \right)=2 \quad &\longrightarrow& r\left ( t  \right)\in \left
\{ -1,1 \right \},     \nonumber \\
 \ x\left ( t  \right)=3 \quad &\longrightarrow& r\left ( t  \right)\in \left
\{ -1,1 \right \},     \nonumber \\
 \ x\left ( t  \right)=4 \quad &\longrightarrow& r\left ( t  \right)\in \left
\{ -2,0,2 \right \}.
\end{eqnarray}
The mapping between transitions $x\left ( t  \right)$ and allowed values of the mid-price changes $r\left ( t  \right)$ has been done by using the cases shown in Fig. \ref{fig:Transition}. This assumption is grounded on the empirical observation that mid-price changes $\left|r\left(t\right)\right|>2$ are extremely rare for large tick assets (see Section 4). 

In the simplest model, we assume that the probability distribution of returns between two transactions depends only on the spread transition between them. We can therefore define the following conditional probabilities defining the process of returns:
\begin{eqnarray}
\label{eq:1.7}
 \ P(r\left ( t  \right)=\pm2|x\left ( t 
\right)=1;\boldsymbol{\theta})&=&\theta_{1},     \nonumber \\
\ P(r\left ( t  \right)=0|x\left ( t 
\right)=1;\boldsymbol{\theta})&=&1-2\theta_{1},     \nonumber \\
 \ P(r\left ( t  \right)=\pm1|x\left ( t 
\right)=2;\boldsymbol{\theta})&=&1/2,     \nonumber \\
 \ P(r\left ( t  \right)=\pm1|x\left ( t 
\right)=3;\boldsymbol{\theta})&=&1/2,     \nonumber \\
 \ P(r\left ( t  \right)=\pm2|x\left ( t 
\right)=4;\boldsymbol{\theta})&=&\theta_{4},     \nonumber \\
\ P(r\left ( t  \right)=0|x\left ( t 
\right)=4;\boldsymbol{\theta})&=&1-2\theta_{4}.      
\end{eqnarray}
Notice that we have assumed symmetric distributions for returns between positive and negative values and
$\boldsymbol{\theta}=( \theta_{1},\theta_{4})'$ is the parameter
vector that we can estimate from data. The parameter $\theta_1$ ($\theta_4$) describes the probability that mid-price changes when the spread remains constant at one (two) ticks. 

The coupled model of spread and return described here will be termed the MS model. When we consider the special case of spread described by a Bernoulli process we will refer to it as the MS$_{B}$ model.

\medskip

\noindent {\it Properties of price returns.} Here we derive the moments and the autocorrelation functions 
$corr\left(r\left(t\right),r\left(t+\tau \right)\right)\equiv\zeta\left(\tau \right)$ and $corr\left(r^{2}\left(t\right),r^{2}\left(t+\tau \right)\right)\equiv\rho\left(\tau \right)$ under the MS model. The quantity $\zeta\left(\tau \right)$ is useful  to study the statistical efficency of price, while $\rho\left(\tau \right)$ describes volatility clustering in transaction time.  

We compute first the  vectors of conditional first, second and fourth moments
\begin{eqnarray}
\label{eq:1.7.1}
\ E\left[r\left(t\right)|x\left(t\right)=k\right]&=&m_{1,k}, \nonumber \\
\ E\left[r^{2}\left(t\right)|x\left(t\right)=k\right]&=&m_{2,k}, \nonumber \\
\ E\left[r^{4}\left(t\right)|x\left(t\right)=k\right]&=&m_{4,k}.
\end{eqnarray}
where $m_{j,k}$ indicates the $k-$th component of the vector $\boldsymbol{m}_{j}$. We have  $\boldsymbol{m}_{1}=\boldsymbol{0}$, $\boldsymbol{m}_{2}=\left( 8\theta_{1},1,1,8\theta_{4}\right)'$ and $\boldsymbol{m}_{4}=\left(32\theta_{1},1,1,32\theta_{4}\right)'$. 
Then we compute unconditional moments by using the stationary vector $\boldsymbol{\lambda}$ as
\begin{eqnarray}
\label{eq:1.7.2}
\ E\left[r\left(t\right)\right]&=&\sum_{k=1}^4 E\left[r\left(t\right)|x\left(t\right)=k\right]P\left[x\left(t\right)=k\right]=\boldsymbol{m}_{1}'\boldsymbol{\lambda}, \nonumber \\
\ E\left[r^{2}\left(t\right)\right]&=&\sum_{k=1}^4 E\left[r^{2}\left(t\right)|x\left(t\right)=k\right]P\left[x\left(t\right)=k\right]=\boldsymbol{m}_{2}'\boldsymbol{\lambda}, \nonumber \\
\ E\left[r^{4}\left(t\right)\right]&=&\sum_{k=1}^4 E\left[r^{4}\left(t\right)|x\left(t\right)=k\right]P\left[x\left(t\right)=k\right]=\boldsymbol{m}_{4}'\boldsymbol{\lambda}, \nonumber \\
\ Var\left[r\left(t\right)\right]&=&\boldsymbol{m}_{2}'\boldsymbol{\lambda}-\left(\boldsymbol{m}_{1}'\boldsymbol{\lambda}\right)^{2},\nonumber \\
\ Var\left[r^{2}\left(t\right)\right]&=&\boldsymbol{m}_{4}'\boldsymbol{\lambda}-\left(\boldsymbol{m}_{2}'\boldsymbol{\lambda}\right)^{2}, 
\end{eqnarray}

In order to compute the linear autocorrelation function  $\zeta(\tau)$  we need to compute $E\left[r\left(t\right)r\left(t+\tau \right)\right]$, by using conditional independence of $r\left(t\right)$ with respect to $x\left(t\right)$. We obtain:
\begin{eqnarray}
\label{eq:1.7.3}
\ &E&\left[r\left(t\right)r\left(t+\tau \right)\right]= \nonumber \\
\ &=& \sum_{i=1}^4 \sum_{j=1}^4 E\left[r\left(t\right)r\left(t+\tau \right)|x\left(t\right)=i,x\left(t+\tau\right)=j\right]P\left[x\left(t\right)=i,x\left(t+\tau \right)=j\right] \nonumber \\
\ &=& \sum_{i=1}^4 \sum_{j=1}^4 E\left[r\left(t\right)|x\left(t\right)=i\right]E\left[r\left(t+\Delta t\right)|x\left(t+\tau \right)=j\right]P\left[x\left(t\right)=i,x\left(t+\tau \right)=j\right] \nonumber \\
\ &=& \sum_{i=1}^4 \sum_{j=1}^4 m_{1,i}m_{1,j}\lambda_{i}M^{\tau }_{ij}=\boldsymbol{\lambda}'\Lambda M^{\tau } \boldsymbol{m}_{1}, \nonumber \\
\end{eqnarray}
where we define the matrix $\Lambda=diag\left(m_{1,1},m_{1,2},m_{1,3},m_{1,4}\right)$. 
The autocorrelation function of returns is given by:
\begin{equation}
\label{eq:1.7.4}
\zeta\left(\tau \right)=\frac{\boldsymbol{\lambda}'\Lambda M^{\tau} \boldsymbol{m}_{1}-\left(\boldsymbol{m}_{1}'\boldsymbol{\lambda}\right)^{2}}{\boldsymbol{m}_{2}'\boldsymbol{\lambda}-\left(\boldsymbol{m}_{1}'\boldsymbol{\lambda}\right)^{2}},
\end{equation}
in our specific case $\zeta\left(\tau \right)=0$ because symmetry leads to $\boldsymbol{m}_{1}=0$. 

We also compute the autocorrelation function of squared returns $\rho(\tau)$ which is equal to
\begin{equation}
\label{eq:1.7.5}
\rho\left(\tau \right)=\frac{\boldsymbol{\lambda}'\Sigma M^{\tau}\boldsymbol{m}_{2}-\left(\boldsymbol{m}_{2}'\boldsymbol{\lambda}\right)^{2}}
{\boldsymbol{m}_{4}'\boldsymbol{\lambda}-\left(\boldsymbol{m}_{2}'\boldsymbol{\lambda}\right)^2},
\end{equation}
where we define the matrix $\Sigma=diag\left(m_{2,1},m_{2,2},m_{2,3},m_{2,4}\right)$. 

As expected, both correlation functions depends on powers of the transition probability matrix $M$. For a Markov process, $M$ is diagonalizable and we can write $M^{\tau}=CM_{D}^{\tau}C^{-1}$, where:
$$
M_{D}^{\tau}=
\begin{pmatrix} 
0 & 0 & 0 & 0 \\
0 & 0 & 0 & 0 \\
0 & 0 & 1 & 0 \\
0 & 0 & 0 & \left(p_{11}-p_{21}\right)^{\tau}  
\end{pmatrix},
\quad C=
\begin{pmatrix}
1 & 0 & 1 & 1 \\
\frac{p_{11}}{\left(p_{11}-1\right)} & 0 & 1 & \frac{p_{21}}{\left(p_{11}-1\right)} \\
0 & 1 & 1 & 1 \\
0 & \frac{p_{21}}{\left(p_{21}-1\right)} & 1 & \frac{p_{21}}{\left(p_{11}-1\right)}
\end{pmatrix}.
$$   

In the limit case in which the spread is described by a Bernoulli process, the matrix $M_{B}$ is not
diagonalizable but has all eigenvalues in $\mathbb{R}$, i.e. $ sp\left(M_{B}\right)=\left(0,0,0,1 \right)$, and we can compute its Jordan canonical form $J_{B}$. Thus we can rewrite the lag dependence as
$M_{B}^{\tau}=EJ_{B}^{\tau}E^{-1}$, where:
$$
J_{B}=
\begin{pmatrix} 
0 & 1 & 0 & 0 \\
0 & 0 & 0 & 0 \\
0 & 0 & 1 & 0 \\
0 & 0 & 0 & 0  
\end{pmatrix},
\quad E=
\begin{pmatrix}
\left(p-p^{2}\right) & \left(1-p^{2}\right) & p^{2} & 0 \\
-p^{2} & -p^{2} & p^{2} & 0 \\
\left(p-p^2\right) & -p^{2} & p^{2} & \frac{p-1}{p} \\
-p^{2} & -p^{2} & p^{2} & 1
\end{pmatrix}.
$$
The structure of block diagonal matrix $J_{B}$ implies that $J_{B}^{\tau}=J_{B}^{2}=0,\ \forall \tau \geq 2$ and that $\rho\left(\tau \right)$ is a constant function for $\tau \geq2$.

\medskip

\begin{figure}
	\centering
\includegraphics[width=0.49\textwidth]{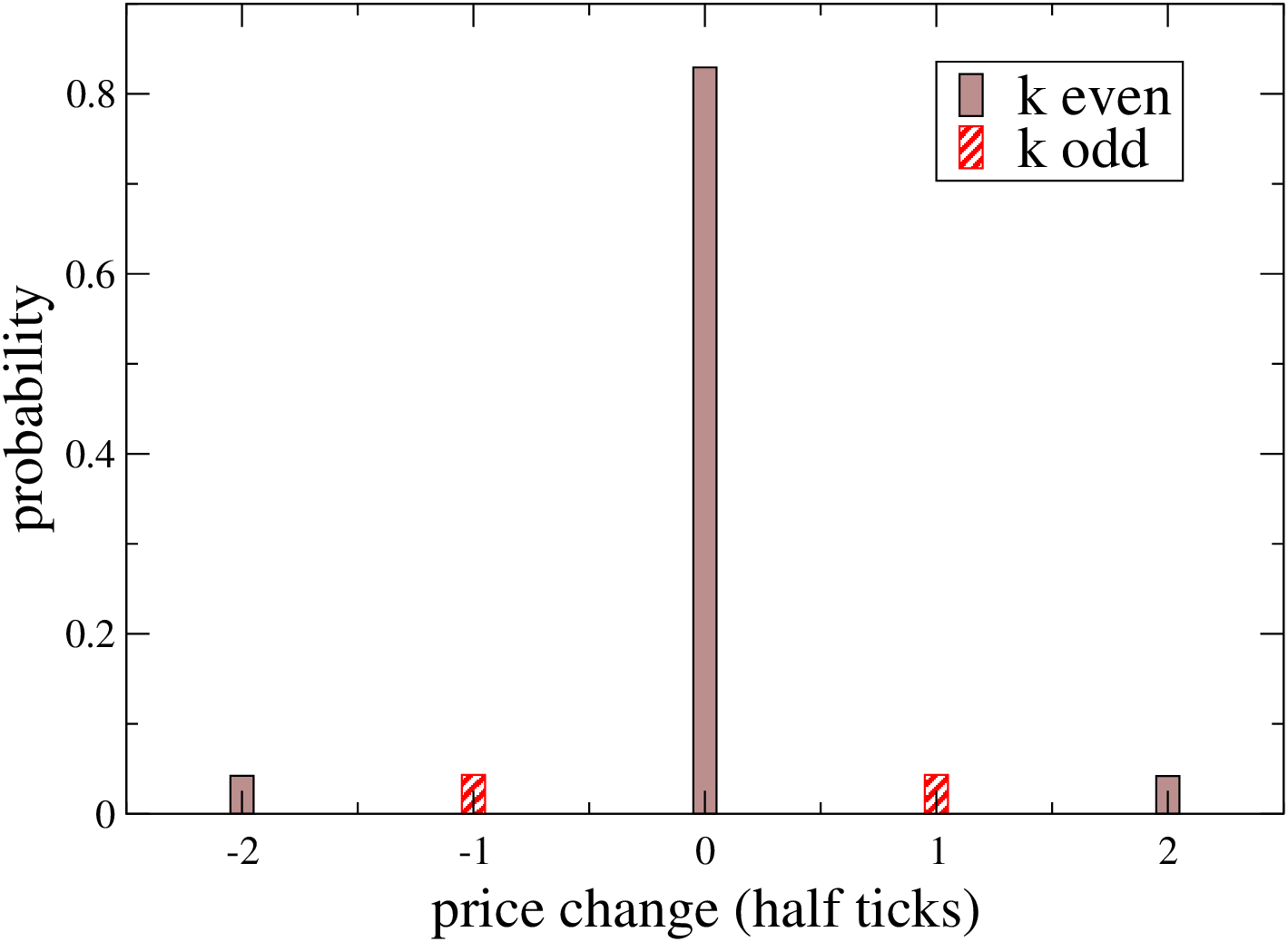}
\includegraphics[width=0.49\textwidth]{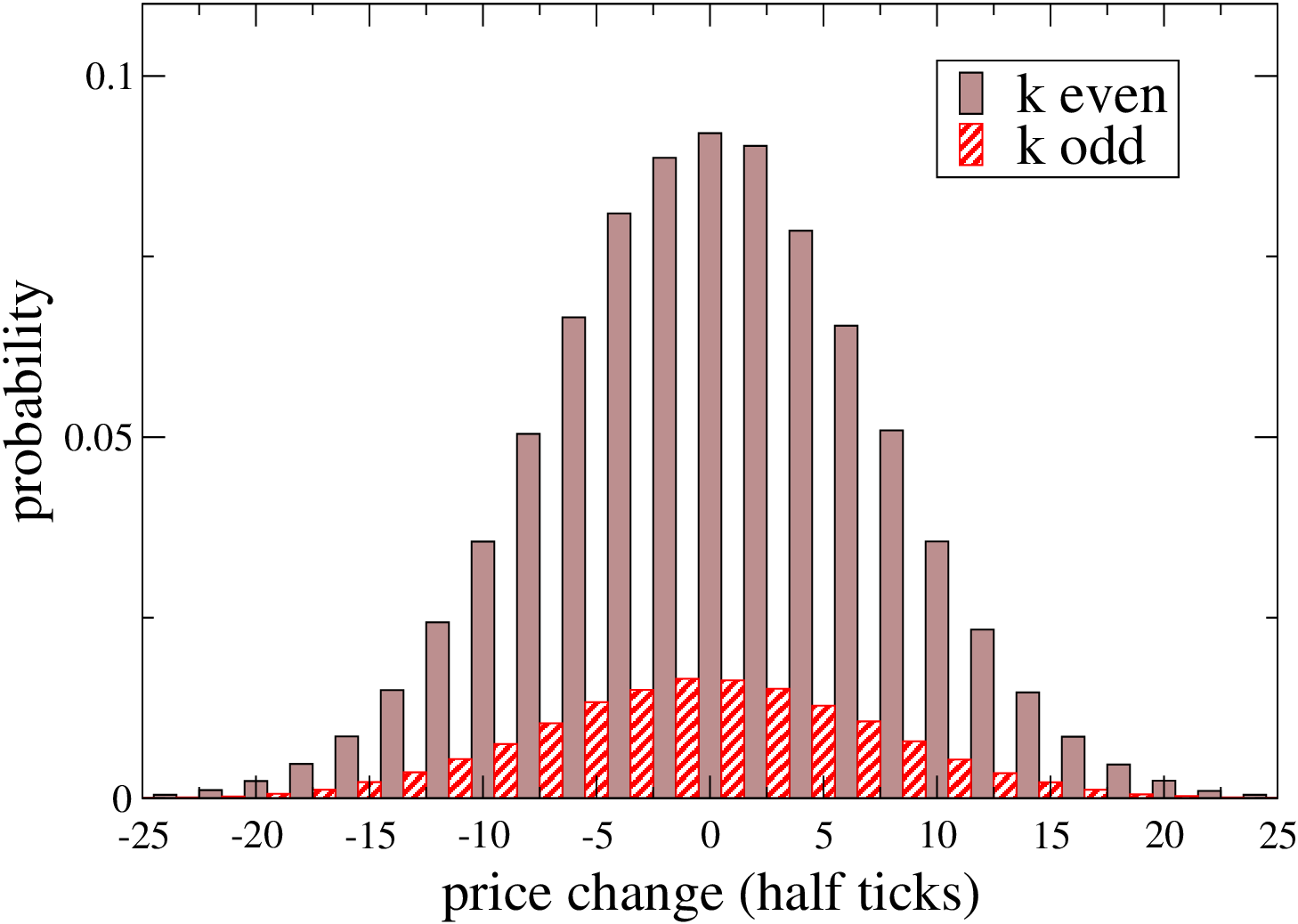}
	\caption{Unconditional distributions of mid-price changes for the simulation of MS model calibrated on  MSFT. The left panel shows $r\left(t,\Delta t=1\right)=p_{m}\left(t+1\right)-p_{m}\left(t\right)$, whereas the right panel shows 
$r\left(t,\Delta t=128\right)=p_{m}\left(t+128\right)-p_{m}\left(t\right)$.}
	\label{fig:MSret}
\end{figure}

{\it Discussion.} The qualitative comparison of real data and model shows that the MS model is able to reproduce the distribution of returns quite well. This can be seen by comparing Fig.\ref{fig:ret} with Fig.\ref{fig:MSret}. It is worth noting that, at least qualitatively, also the Bernoulli model MS$_B$ is able to reproduce the underestimation of odd values of returns with respect to the even values, as observed in real data. Therefore it is the coupling of spread and return, rather than the memory properties of spread, which is responsible of the behavior of the aggregated return distribution of Fig. \ref{fig:ret}.
It is also possible to show that the model has linearly uncorrelated returns, as observed in real data, at least for lags larger than few transactions.

However the model fails to describe the volatility clustering. In fact, we can prove that $\rho\left(\tau\right)$ is an exponential function, $\exp(-a\tau$, with $a=-\ln(p_{11}-p_{21})$, i.e. the model describes an exponentially decaying volatility clustering. As the data calibration shows (see Section 5 and Figure 5), the predicted behavior of $\rho\left(\tau\right)$ under the MS model is much smaller than the one observed in real data. Therefore this model is unable to reproduce the volatility clustering as well as any long memory property. This observation motivates us to develop a model that, preserving the structure of the coupling between spread and returns discussed so far, is able to describe non exponential volatility clustering. This model is developed in the next section.

\subsection{A double chain Markov model with logit regression}
\label{sec:A double chain Markov model with logit regression}

The Markov switching model is not able to explain the empirically observed correlation of squared returns shown in Fig. \ref{fig:volclust}. Therefore in the second class of models we consider an autoregressive switching model for returns \citep{Ferland,Schnatter}
in order to study correlation of squared returns. 
The idea is to use  logit regressions on past values of variables, i.e. returns and squared returns in order to reduce the number of parameters that one would have with an higher order Markov process. 
The model is thus defined
by the following conditional probabilities \citep{Fokianos}:
\begin{eqnarray}
\label{eq:1.8}
\ P\left ( r\left ( t \right )|x\left ( t \right )=k,\boldsymbol{\Omega}\left
( t-1 \right );\boldsymbol{\theta}_{k} \right ),\quad k \in \left \{ 1,2,3,4
\right \} \nonumber \\
\ \boldsymbol{\Omega}'\left ( t-1 \right )=\left ( r^{2}\left ( t-1 \right
),...,r^{2}\left ( t-p \right ),r\left ( t-1 \right ),...,r\left ( t-e \right )
\right
)=\left(\boldsymbol{\Omega}_{r^{2}}',\boldsymbol{\Omega}_{r}'\right)
\nonumber \\
\ \boldsymbol{\theta}_{k}'=\left
(\alpha_{k},\boldsymbol{\beta}_{k}',\boldsymbol{\gamma}_{k}'  \right ),
\end{eqnarray}
where we define an informative $(p+e)$-dimensional vector of regressors
$\boldsymbol{\Omega}$, made of the past $e$ returns and $p$ squared returns. Each parameter vector $\boldsymbol{\theta}_{k}$ is composed by the scalar 
$\alpha_{k}$, the $p$-dimensional vector $\boldsymbol{\beta}_{k}$
which describes the regression on past values of squared returns, and the $e$-dimensional
vector $\boldsymbol{\gamma}_{k}$ which  describes the regression on past returns. 

In order to handle the discreteness of returns we make use of a logit regression. To this end we first convert the returns series in
a binary series $b\left ( t \right )\in \left \{ 0,1 \right \}$. When the spread remains constant between $t$ and $t+1$ (i.e. $x(t)=1$ or $x(t)=4$),  we set
\begin{eqnarray}
\label{eq:1.9}
\ r\left ( t \right )=\pm2\ &\longrightarrow&\ b\left ( t \right )=1 \nonumber \\
\ r\left ( t \right )=0\ &\longrightarrow&\ b\left ( t \right )=0
\end{eqnarray}
while when the spread changes, (i.e. $x(t)=2$ or $x(t)=3$) we set
\begin{eqnarray}
\label{eq:1.9b}
\ r\left ( t \right )=1\ &\longrightarrow&\ b\left ( t \right )=1 \nonumber \\
\ r\left ( t \right )=-1\ &\longrightarrow&\ b\left ( t \right )=0
\end{eqnarray}
Then by denoting by $\eta_{k}\left ( t \right )$ 
the conditional probability of having $b\left ( t \right )=1$, the logit regression is
\begin{eqnarray}
\label{eq:1.10}
\ P\left ( b\left ( t \right )|x\left ( t \right )=k,\boldsymbol{\Omega}\left
( t-1 \right );\boldsymbol{\theta}_{k} \right )&=&\exp\left \{ b\left ( t \right
)\log\left ( \frac{\eta_{k}\left ( t \right )}{1-\eta_{k}\left ( t \right )}
\right )+\log\left(1-\eta_{k}\left ( t \right )\right) \right \}  \nonumber \\
\ \eta_{k}\left ( t \right
)&=&\frac{\exp\left(\alpha_{k}+\boldsymbol{\Omega}_{r^{2}}'\left ( t-1 \right
)\boldsymbol{\beta}_{k}+\boldsymbol{\Omega}_{r}'\left ( t-1 \right
)\boldsymbol{\gamma}_{k}\right)}{1+\exp\left(\alpha_{k}+\boldsymbol{\Omega}_{r^{
2}}'\left ( t-1 \right
)\boldsymbol{\beta}_{k}+\boldsymbol{\Omega}_{r}'\left ( t-1 \right
)\boldsymbol{\gamma}_{k}\right)}  \nonumber \\ 
\end{eqnarray} 
and we finally obtain the process for $r\left(t\right)$ by:
\begin{eqnarray}
\label{eq:1.11}
\left\{\begin{array}{c}
\ P\left(r\left(t\right)=\pm2|x\left(t\right)=1,\boldsymbol{\Omega}\left (
t-1 \right );\boldsymbol{\theta}_{1}\right)=\eta_{1}\left(t\right)/2, \\
\ P\left(r\left(t\right)=0|x\left(t\right)=1,\boldsymbol{\Omega}\left ( t-1
\right );\boldsymbol{\theta}_{1}\right)=1-\eta_{1}\left(t\right)
\end{array}\right. \nonumber \\
\left\{\begin{array}{c}
\ P\left(r\left(t\right)=1|x\left(t\right)=2,\boldsymbol{\Omega}\left ( t-1
\right );\boldsymbol{\theta}_{2}\right)=\eta_{2}\left(t\right),  \\
\ P\left(r\left(t\right)=-1|x\left(t\right)=2,\boldsymbol{\Omega}\left ( t-1
\right );\boldsymbol{\theta}_{2}\right)=1-\eta_{2}\left(t\right), 
\end{array}\right. \nonumber \\
\left\{\begin{array}{c}
\ P\left(r\left(t\right)=1|x\left(t\right)=3,\boldsymbol{\Omega}\left ( t-1
\right );\boldsymbol{\theta}_{3}\right)=\eta_{3}\left(t\right),  \\
\ P\left(r\left(t\right)=-1|x\left(t\right)=3,\boldsymbol{\Omega}\left ( t-1
\right );\boldsymbol{\theta}_{3}\right)=1-\eta_{3}\left(t\right).
\end{array}\right. \nonumber \\
\left\{\begin{array}{c}
\ P\left(r\left(t\right)=\pm2|x\left(t\right)=4,\boldsymbol{\Omega}\left (
t-1 \right );\boldsymbol{\theta}_{4}\right)=\eta_{4}\left(t\right)/2,\\
\ P\left(r\left(t\right)=0|x\left(t\right)=4,\boldsymbol{\Omega}\left ( t-1
\right );\boldsymbol{\theta}_{4}\right)=1-\eta_{4}\left(t\right), 
\end{array}\right. 
\end{eqnarray} 

These equations define the general DCMM($e,p$) model. In the rest of the paper we will consider the case $e=0$ and for the sake of simplicity we will denote DCMM($p$)=DCMM($0,p$).
In our case the independent latent Markov process is represented by the transition process $x\left(t\right)$ and the dependent Markov process is represented by the
$r\left(t\right)$  processes. The form of stochastic dependence is defined by the logit rules in Eq. (\ref{eq:1.11}).

For the sake of clarity, here we consider the case $p=1$, while its extension to a general value for $p$ is considered in Appendix \ref{sec:Appendix A}. 
The definition of the process for $r\left(t\right)\in \left \{ -2,-1,0,1,2 \right \}$, and $i,j \in \left \{ 1,2,3,4,5 \right \}$, in the case of $p=1$ (DCMM(1)) is the following:
\begin{equation}
\label{eq:2.1}
P\left(r\left(t\right)=\left(3-j\right)|x\left(t\right)=k, r\left(t-1\right)=\left(3-i\right);\boldsymbol{\theta}_{k} \right)=A_{k,ij}\ . 
\end{equation}
We have four possible transition matrices $A_{x\left(t\right)=k}$ for $k \in \left \{ 1,2,3,4 \right \}$, determined by the latent process $x\left(t\right)$:
$$
A_{x\left(t\right)=1}=
\begin{pmatrix}
\eta_{1}\left(r^{2}\left(t-1\right)=4\right)/2 & 0 & 1-\eta_{1}\left(r^{2}=4\right) & 0 & \eta_{1}\left(r^{2}=4\right)/2 \\
\eta_{1}\left(r^{2}\left(t-1\right)=1\right)/2 & 0 & 1-\eta_{1}\left(r^{2}=1\right) & 0 & \eta_{1}\left(r^{2}=1\right)/2 \\
\eta_{1}\left(r^{2}\left(t-1\right)=0\right)/2 & 0 & 1-\eta_{1}\left(r^{2}=0\right) & 0 & \eta_{1}\left(r^{2}=0\right)/2 \\
\eta_{1}\left(r^{2}\left(t-1\right)=1\right)/2 & 0 & 1-\eta_{1}\left(r^{2}=1\right) & 0 & \eta_{1}\left(r^{2}=1\right)/2 \\
\eta_{1}\left(r^{2}\left(t-1\right)=4\right)/2 & 0 & 1-\eta_{1}\left(r^{2}=4\right) & 0 & \eta_{1}\left(r^{2}=4\right)/2  
\end{pmatrix}
$$
$$
A_{x\left(t\right)=2}=
\begin{pmatrix}
0 & \eta_{2}\left(r^{2}\left(t-1\right)=4\right) & 0 & 1-\eta_{2}\left(r^{2}=4\right) & 0 \\
0 & \eta_{2}\left(r^{2}\left(t-1\right)=1\right) & 0 & 1-\eta_{2}\left(r^{2}=1\right) & 0 \\
0 & \eta_{2}\left(r^{2}\left(t-1\right)=0\right) & 0 & 1-\eta_{2}\left(r^{2}=0\right) & 0 \\
0 & \eta_{2}\left(r^{2}\left(t-1\right)=1\right) & 0 & 1-\eta_{2}\left(r^{2}=1\right) & 0 \\
0 & \eta_{2}\left(r^{2}\left(t-1\right)=4\right) & 0 & 1-\eta_{2}\left(r^{2}=4\right) & 0  
\end{pmatrix}
$$
where we have specified the temporal dependence in regressors only in the first column. The others two matrices have same definitions: 
$A_{4}=A_{1}\left(\eta_{1}\rightarrow \eta_{4}\right)$ and $A_{3}=A_{2}\left(\eta_{2}\rightarrow \eta_{3}\right)$. In this way, assuming
that the latent process has reached the stationary distribution defined by Eq. \ref{eq:1.5}, we can define an overall Markov chain by the transition matrix $N$ that describes the $r\left(t\right)$ process:
\begin{equation}
\label{eq:2.2}
N=\sum_{k=1}^4 \lambda_{k} A_{k}.
\end{equation}
The matrix $N$ is defined by $6+4p$ parameters: $p_{11}, p_{21}, \alpha_{k}, \boldsymbol{\beta}_{k}'$.

The probabilities of the process for $r^2\left(t\right)\in \left \{ 0,1,4 \right \}$, and $i,j \in \left \{ 1,2,3 \right \}$, in the case of $p=1$ (DCMM(1)) is
\begin{equation}
\label{eq:2.3}
P\left(r^2\left(t\right)=\left(3-j\right)^2|x\left(t\right)=k, r^2\left(t-1\right)=\left(3-i\right)^2;\boldsymbol{\theta}_{k} \right)=V_{x\left(t\right),ij}\ . 
\end{equation}
which can be calculated from the knowledge of the matrix $A$. 

In particular, we have four possible transition matrices $V_{x\left(t\right)=k}$ for $k \in \left \{ 1,2,3,4 \right \}$, determined by the latent process $x\left(t\right)$:
$$
V_{x\left(t\right)=1}=
\begin{pmatrix}
\eta_{1}\left(r^2\left(t-1\right)=4\right) & 0 & 1-\eta_{1}\left(r^2=4\right) \\
\eta_{1}\left(r^2\left(t-1\right)=1\right) & 0 & 1-\eta_{1}\left(r^2=1\right) \\ 
\eta_{1}\left(r^2\left(t-1\right)=0\right) & 0 & 1-\eta_{1}\left(r^2=0\right) 
\end{pmatrix},
\quad V_{x\left(t\right)=2}=
\begin{pmatrix}
 0 & 1 & 0\\
 0 & 1 & 0\\
 0 & 1 & 0
\end{pmatrix}.
$$
We can define an overall Markov process for $r^2\left(t\right)$ described by a transition matrix $S$, assuming that the transition process $x\left(t\right)$ has reached the stationary distribution:
\begin{equation}
\label{eq:2.4}
S=\sum_{k=1}^4 \lambda_{k} V_{k}.
\end{equation}
The matrix $S$ is defined by $4+2p$ parameters: $p_{11}, p_{21}, \alpha_{k}, \boldsymbol{\beta}_{k}'$, where $k \in \left \{ 1,4\right \}$. 
The function $corr\left(r^{2}\left(t\right),r^{2}\left(t+\tau \right)\right)=\rho\left(\tau\right)$ for the DCMM(1) process is the correlation
of the Markov(1) process defined by $S$.
We solve the eigenvalue equation for $S$ relative to the eigenvalue $1$ in order to determine the stationary probability vector $\boldsymbol{\psi}$:
\begin{equation}
\label{eq:2.5}
S'\boldsymbol{\psi}= \boldsymbol{\psi},
\end{equation}
the entire spectrum is given by $sp\left(S\right)=\left(0,1,e_{3}\right)$, where the last eigenvalue is:
\begin{equation}
\label{eq:2.6}
e_{3}=\frac{-\left[\left(\eta_{4}\left(0\right)-\eta_{4}\left(4\right)\right)\left(1-p_{11}-p_{21}+p_{11}p_{21}\right)+\left(\eta_{1}\left(0\right)-\eta_{1}\left(4\right)\right)p_{11}p_{21}\right]}{p_{21}-p_{11}+1}.
\end{equation}
If we define the vectors $\boldsymbol{\delta}$, $\boldsymbol{\delta}_{2}$ and $\boldsymbol{\xi}$, where $\delta_{i}=\left(3-i\right)^2$, $\delta_{2,i}=\left(3-i\right)^4$ and $\boldsymbol{\xi}=\boldsymbol{\delta}\odot\boldsymbol{\psi}$ , the moments are given by:
\begin{eqnarray}
\label{eq:2.7}
\ E\left[r^{2}\left(t\right)\right]&=& \boldsymbol{\delta}'\boldsymbol{\psi}, \nonumber \\
\ E\left[r^{4}\left(t\right)\right]&=& \boldsymbol{\delta}_{2}'\boldsymbol{\psi}, \nonumber \\
\ E\left[r^{2}\left(t\right)r^{2}\left(t+\tau\right)\right]&=& \boldsymbol{\xi}'S^{\tau}\boldsymbol{\delta}.
\end{eqnarray}
Finally, we have the expression for $\rho\left(\tau\right)$ in the case $p=1$:
\begin{equation}
\label{eq:2.8}
\rho\left(\tau\right)=\frac{\boldsymbol{\xi}'S^{\tau}\boldsymbol{\delta}-\left(\boldsymbol{\delta}'\boldsymbol{\psi}\right)^{2}}{\boldsymbol{\delta}_{2}'\boldsymbol{\psi}-\left(\boldsymbol{\delta}'\boldsymbol{\psi}\right)^{2}}.
\end{equation}
The generalization of the calculation of $\rho\left(t\right)$ to any value of the order $p$ is reported in the appendix \ref{sec:Appendix A}.

In order to estimate the parameter vector
$\boldsymbol{\theta}'=\left(\boldsymbol{\theta}_{1}',\boldsymbol{\theta}_{
2}',\boldsymbol{\theta}_{3}',\boldsymbol{\theta}_{4}'\right)$ 
we maximize the  partial-loglikelihood, 
\begin{equation}
\label{eq:1.12new}
\mathcal{L}\left(\boldsymbol{\theta}\right)=\sum_{t=p+1}^T\log\left[\sum_{k=1}^4
P\left(x\left(t\right)=k|\boldsymbol{\Omega}\left ( t-1 \right );\boldsymbol{\theta}_{k}\right)P\left ( b\left ( t \right )|x\left ( t
\right )=k,\boldsymbol{\Omega}\left ( t-1 \right );\boldsymbol{\theta}_{k}
\right )\right],
\end{equation}
where  $T$ is the length of sample, and we assume
that parameters $p_{11}$ and $p_{21}$ are known.
Since the dynamics of spread transitions is independent from the past informative set, i.e. $P\left(x\left(t\right)=k|\boldsymbol{\Omega}\left ( t-1 \right );\boldsymbol{\theta}_{k}\right)=P\left(x\left(t\right)=k\right)$,
 we have
\begin{equation}
\label{eq:1.12}
\mathcal{L}\left(\boldsymbol{\theta}\right)=\sum_{t=p+1}^T\log\left[\sum_{k=1}^4
P\left(x\left(t\right)=k\right)P\left ( b\left ( t \right )|x\left ( t
\right )=k,\boldsymbol{\Omega}\left ( t-1 \right );\boldsymbol{\theta}_{k}
\right )\right],
\end{equation}
In the case of large tick assets, it is $\lambda_{1}\approx 1$ and we can use the approximation
\begin{equation}
\label{eq:1.13}
\mathcal{L}\left(\boldsymbol{\theta}\right)\approx \sum_{t=p+1}^T\log \biggl( P\left (
b\left ( t \right )|x\left ( t \right )=1,\boldsymbol{\Omega}\left ( t-1
\right );\boldsymbol{\theta}_{1} \right )\biggr).
\end{equation}
For example for MSFT we have $\lambda_{1}\approx0.9$.
With this approximation we estimate only the vector $\boldsymbol{\theta}_{1}$ and the parameter
$\theta_{4}$ of Eq.\ref{eq:1.7}, that are enough in order to define matrices $V_{k}$.  Moreover we can approximate 
$$
V_{x\left(t\right)=4}\approx
\begin{pmatrix}
2\theta_{4} & 0 & 1-2\theta_{4}\\
2\theta_{4} & 0 & 1-2\theta_{4}\\
2\theta_{4} & 0 & 1-2\theta_{4}. 
\end{pmatrix}
$$
In this way we neglect the contribution of regressors $\boldsymbol{\Omega}\left(t-1\right)$ (weighted by $\boldsymbol{\beta}_{4}$) and make use of the simpler expression in Eq. \ref{eq:1.7} when $x\left(t\right)=4$. As before, this approximation holds if the weight of $V_{x\left(t\right)=4}$ is negligible, i.e. $\lambda_{4}\approx0$, i.e. when there is a small
number of spread transitions $s\left(t\right)=2\rightarrow s\left(t+1\right)=2$. This is the case when we have large tick assets, where we have almost always $s\left(t\right)=1$. In the case of MSFT asset for example we have $\lambda_{4} \simeq 0.04$.

\begin{figure}
	\centering
		\includegraphics[width=0.8\textwidth]{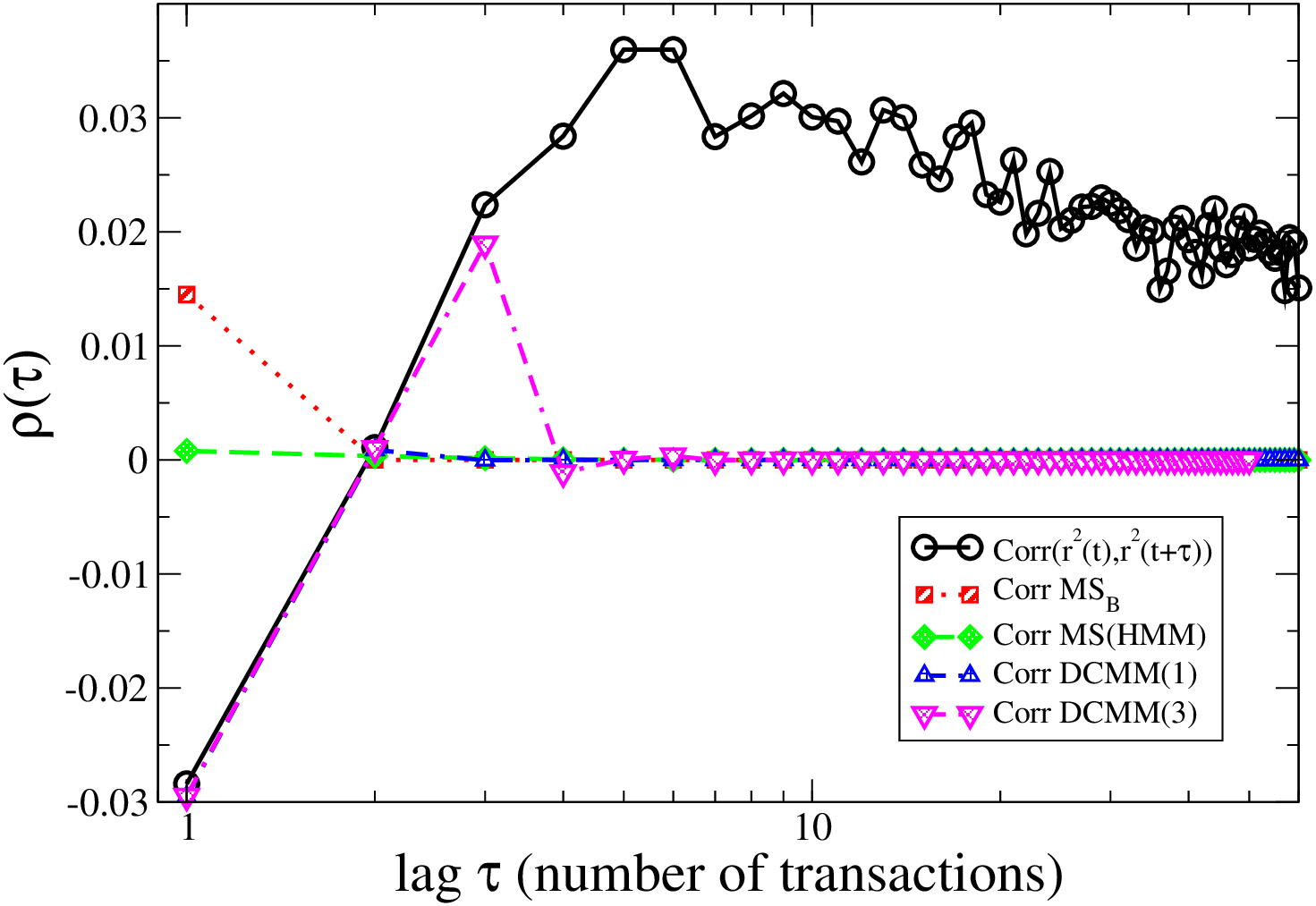}
	\caption{Autocorrelation function of squared returns, $\rho\left(\tau\right)$. The black circles are the real data of MSFT asset.
The red squares are the result of the MS$_{B}$ model, the green diamonds refer to the MS model, the blu up triangles refer to the DCMM(1) model and the pink down triangles refer to DCMM(3) model, all calibrated on the MSFT asset.}
	\label{fig:DCMMlogit}
\end{figure}

We have performed the calculation of the autocorrelation $\rho(\tau)$ of the squared returns for $p=1,3$ and the result is reported in Fig. \ref{fig:DCMMlogit}. We have calibrated the parameters on the MSFT asset (see next Sections for details).
We note that the MS and MS$_B$ models underestimate very strongly $\rho(\tau)$. Note that for the MS model, $\rho(\tau)$ calibrated on real data is very small but not zero as predicted by the theory. The DCMM(p) model, on the other hand, is able to fit very well $\rho(\tau)$ up to lag $\tau=p$. Remarkably the model captures very well also the negative correlation for very short lags. However this observation indicates that an higher order DCMM(p) model might be able to fit better the real data. In the next Sections we will show that this is indeed the case.

\section{Data}
\label{sec:Data and stylized facts}

We have investigated two stocks, namely Microsoft (MSFT) and Cisco (CSCO), both traded at NASDAQ market in the period
July-August 2009, corresponding to 42 trading days. 
Data contains time stamps corresponding to order executions, prices, size of trading volume and direction
of trading. The time resolution is one millisecond. In this article we report mostly the results for MSFT asset, which are very similar to those for CSCO.

\begin{figure}
	\centering
		\includegraphics[width=0.8\textwidth]{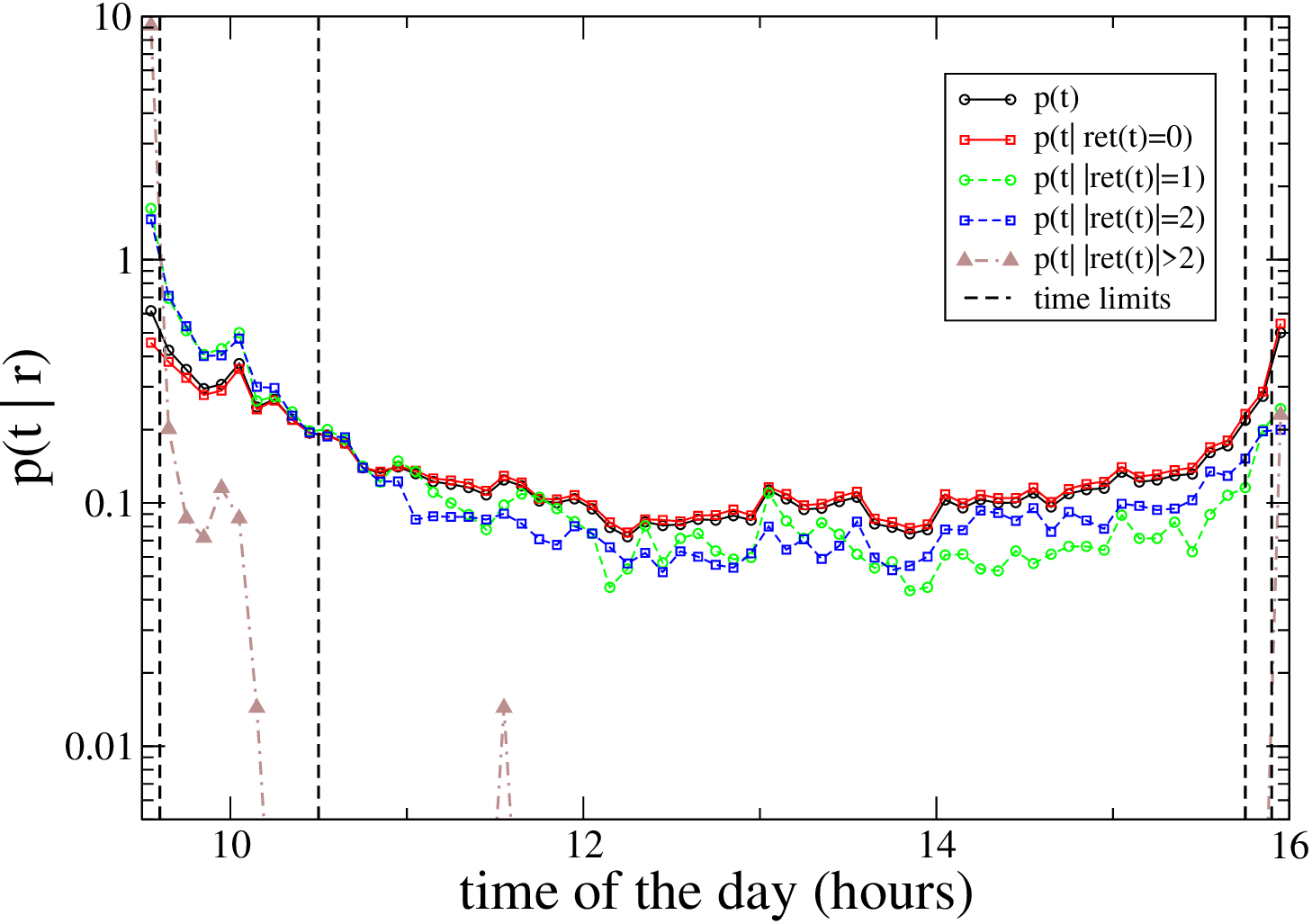}
	\caption{Unconditional and conditional probability distribution of the time of the day when a transaction occurs. 
We bin data into $6$ minute intervals.
}
	\label{fig:Tradeshape}
\end{figure}

Non stationarities can be very important when investigating intraday financial data. For this reason and in order to restrict our empirical analysis to roughly stationary time intervals, we first compute the intensity of trading activity at time $t$ conditional to a specific value $k$ of mid-price change, i.e. $p\left(t|r(t)=k\right)$. 
As we can see from Figure \ref{fig:Tradeshape}, the unconditional trading intensity $p(t)$ is
not stationary during the day \citep{Andersen}. As usual, trading activity is very high at the beginning and at the end of the day. For this reason, we discard transaction data in the first and last six minutes of trading day. Moreover figure shows that the relative frequencies of the three values of returns change during the day, except for returns larger than two ticks that are very rare throughout the day. Most important, in the first part of the day, one tick or two tick returns are more frequent than zero returns, while after approximately $10:30$ the opposite is true. For this reason we split our times series in two subsamples. The
first sample, corresponding to a period of high trading intensity, covers the
time sets $t \in \left(9:36,10:30\right) \cup \left(15:45,15:54\right)$, where time is measured in hours. The
second sample, corresponding to low trading intensity, covers the time set
$t\in \left[10:30,15:45\right]$.
Table 1 reports a summary statistics of the two subsamples.

\begin{table}
\footnotesize 
\centering
\begin{tabular}{|l||l|l|l|l|l||l||}
  \hline
  asset & activity & $\#$ trades  & mean (ticks/2)  & $\sigma$ (ticks/2)  & ex. kurt & $\hat \pi_1$ \\
  \hline
  MSFT  & high & $184,542$ & $-2.82*10^{-4}$ & $0.652$ & $5.13$ & $0.92$ \\ \cline{2-7}
  $42$ days & low  & $348,253$ & $8.96*10^{-4}$ & $0.514$  & $9.89$ & $0.95$ \\
  \hline \hline
  CSCO  & high & $145,084$ & $-1.32*10^{-3}$ & $0.673$ & $4.73$ & $0.92$ \\ \cline{2-7}
  $42$ days & low  & $275,879$ & $1.44*10^{-3}$ & $0.551$  & $8.46$ & $0.95$ \\
  \hline
\end{tabular}
\caption{Summary statistics for assets MSFT and CSCO in the two subsamples of high and low trading activity. $\sigma$ is the standard deviation and ex. kurt is the excess kurtosis of tick by tick returns, and $\hat \pi_1$ is the fraction of time the spread is equal to one tick.}
\end{table}

\begin{figure}
	\centering
		\includegraphics[width=0.8\textwidth]{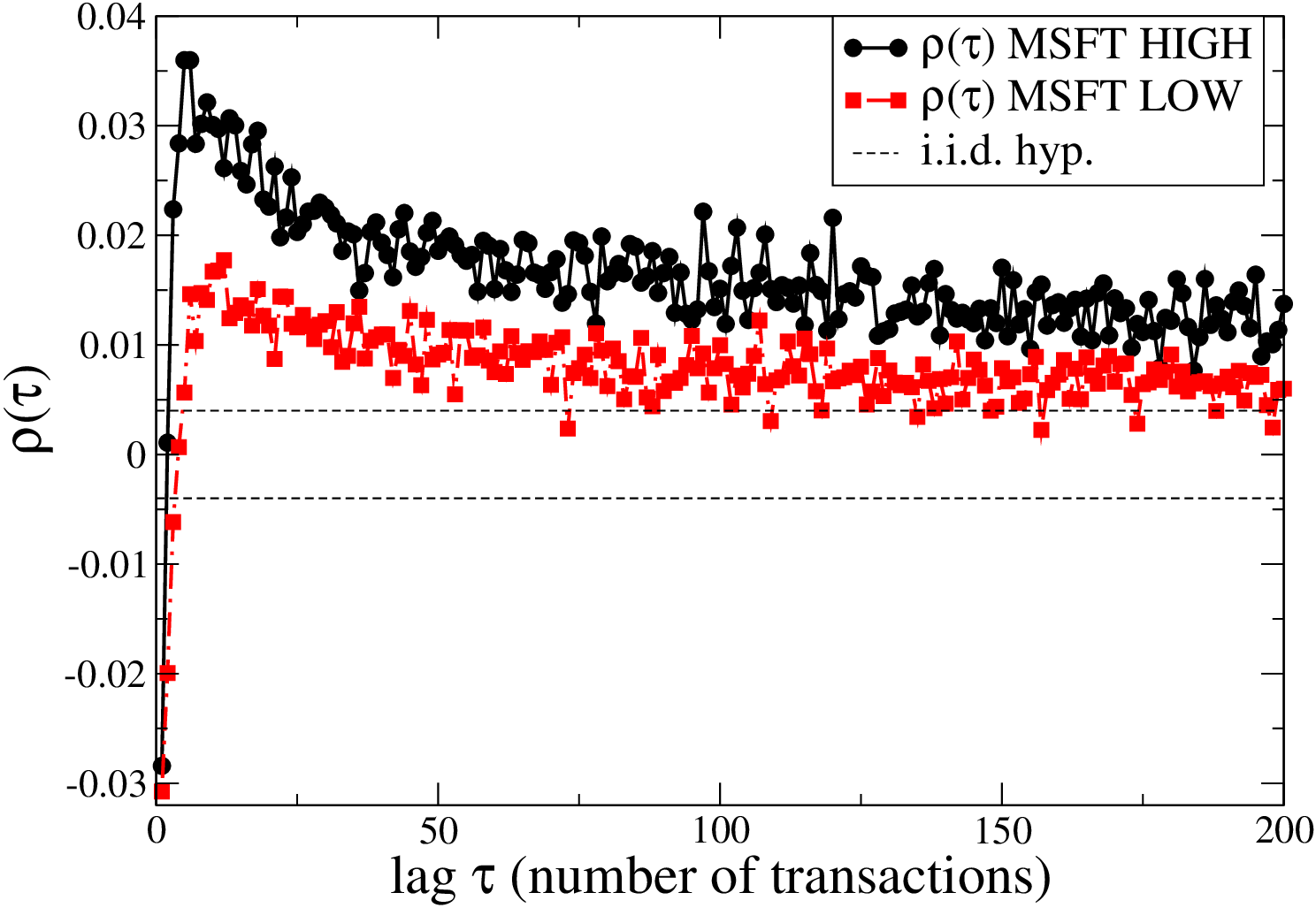} 
	\caption{Sample autocorrelation function of squared returns, $\rho\left(\tau \right)$ for MSFT. Black circles refer to high trading activity series and the red squares refer to
low trading activity series. The 
dashed lines indicate $2\sigma$ confidence intervals in the hypothesis of i.i.d. time series.}
	\label{fig:ACFr2}
\end{figure}

We then analyze the empirical autocorrelation function of squared returns
$corr\left(r^2\left(t\right),r^2\left(t+ \tau \right)\right)=\rho\left(\tau \right)$ for these two series.
As we can see from Fig. \ref{fig:ACFr2}, for $\tau >5$ both time series display a significant
positive and slowly decaying autocorrelation, which is
a quantitative manifestation of volatility
clustering. The series corresponding to low trading
activity displays smaller, yet very persistent, volatility clustering. 

\section{Estimation of the models and comparison with real data}
\label{sec:Simulation}

We have estimated the models described in Secs.\ref{sec:Markov-Switching models} and \ref{sec:A double chain Markov model with logit regression} and we have used Monte Carlo simulations to generate artificial time series calibrated on real data. The properties of these time series have been compared with those from real data.  

More specifically we have considered three models: (i) the MS$_{B}$ model, where spread is described by a Bernoulli process and there are no logit regressors; (ii) the MS model, where spread is a Markov(1) process and there are no logit regressors; (iii) the DCMM($p$) model, where spread is a Markov(1) process and the set of logit regressors includes only the past $p$ values of squared returns. Notice therefore that in this last model we set $e=0$. Finally, we have estimated the model separately for high and low activity regime.


\subsection{Estimation of the models}

From spread and returns data we computed the estimators  $\hat{\pi}_1,\ \hat{p}_{11},\ \hat{p}_{21},\ \hat{\theta}_{1},\ \hat{\theta}_{4}$
of the parameters defined in Sec.\ref{sec:Markov-Switching models}. They are given by
\begin{eqnarray}
\label{eq:1.15}
\ \hat{\pi}_1&=&\frac{n_{1}}{N_{s}}, \nonumber \\
\ \hat{p}_{ij}&=&\frac{n_{ij}}{\sum_{j=1}^2 n_{ij}}, \nonumber \\
\ \hat{\theta}_{k}&=&\frac{1}{2}\left(1-\frac{n_{0k}}{N_{k}}\right),
\end{eqnarray}
where $n_{1}$ is the number of times $s\left(t\right)=1$, $N_{s}$ is the length of the spread time series, $n_{ij}$ is the number of times the value of spread $i$ is followed by the value $j$, $n_{0k}$ is the number of times returns
are zero in the regime $x\left(t\right)=k$, and $N_{k}$ is the length of the subseries of returns in the same regime. For the last estimator $\hat{\theta}_{k}$ we count only zero returns because we assumed that the returns are
distributed symmetrically in the set $\left(-2,0,2\right)$. We have checked that this assumption represents a good approximation for our data sets. The estimated parameters for MSFT asset are shown in Table 2.

\begin{table}
\small 
\centering
\begin{tabular}{ |l|l|l|l|l|l| }
  \hline
  activity & $\hat{\pi}_1$  & $\hat{p}_{11}$ & $\hat{p}_{21}$ & $\hat{\theta}_{1}$ & $\hat{\theta}_{4}$ \\
  \hline
  high  & $9.17*10^{-1}$ & $9.53*10^{-1}$ & $5.22*10^{-1}$  & $4.81*10^{-2}$ & $1.51*10^{-3}$ \\
  \hline
  low  & $9.52*10^{-1}$ & $9.72*10^{-1}$ & $5.50*10^{-1}$  & $2.85*10^{-2}$ & $2.65*10^{-4}$ \\
  \hline 
\end{tabular}
\caption{Estimated parameters for the MSFT asset.}
\end{table}

In order to estimate the DCMM($p$) model we need to estimate the vector $\boldsymbol{\theta}$. For both regimes we use the approximated log-likelihood of Eq. \ref{eq:1.13} because we have for low volatility series $P\left(x\left(t\right)=1\right)\approx0.92$ and for high volatility
$P\left(x\left(t\right)=1\right)\approx0.87$. Thus we need to estimate only the vector $\boldsymbol{\theta}_{1}=\left(\alpha_{1},\boldsymbol{\beta_{1}}'\right)$
by a standard generalized linear regression and we use an iterative reweighted least squares technique \citep{Fokianos}. In this way we generate the returns series in regime $x\left(t\right)=1$, instead for the other regimes the generator
follows the rules in Eq. \ref{eq:1.7}, i.e. we use the estimator $\hat{\theta}_{4}$.
The order of model is fixed to $p=50$ in order to investigate the impact of past squared returns on the returns process. For simplicity we report here only the results from high activity time series.

We find $\alpha_{1}=-2.921(0.019)$ and we report the first $25$ values of $\beta_{1i}$ in Table 3. The estimates of $\beta_{1i}$ are significantly positive for $i>2$ up to $i=50$, with the exception of $i=36,37$. Moreover they display a maximum for $i=6$. We perform a power law fit on these parameters, $\hat{\beta}_{1i}\propto i^{-\alpha}$, and we find a significant exponent $\alpha=0.626(0.068)$. We hypothesize that this functional dependence of $\beta_{1i}$ from $i$ could be connected to the slow decay of the autocorrelation function of squared returns, but we have not investigated further this aspect.

\begin{table}[t]
\footnotesize 
\centering
\begin{tabular}{ |l|l|l|l|l|l| }
  \hline
  $i$ & $\hat{\beta}_{1i}$  & st.error & $z-value$ \\
  \hline
  1  & $-1.56*10^{-1}$ & $9*10^{-3}$ & $-18.4\ ***$ \\
  \hline
  2  & $-4.03*10^{-2}$ & $7.4*10^{-3}$ & $-5.45\ ***$ \\
  \hline 
  3  & $2.18*10^{-2}$ & $7.0*10^{-3}$ & $3.12\ **$ \\
  \hline
  4  & $4.58*10^{-2}$ & $6.9*10^{-3}$ & $6.61\ ***$ \\
  \hline
  5  & $7.13*10^{-2}$ & $6.8*10^{-3}$ & $10.5\ ***$ \\
  \hline
  6  & $7.59*10^{-2}$ & $6.8*10^{-3}$ & $11.2\ ***$ \\
  \hline
  7  & $5.94*10^{-2}$ & $6.9*10^{-3}$ & $8.57\ ***$ \\
  \hline
  8  & $6.06*10^{-2}$ & $6.9*10^{-3}$ & $8.76\ ***$ \\
  \hline
  9  & $5.94*10^{-2}$ & $6.9*10^{-3}$ & $8.55\ ***$ \\
  \hline
 10  & $5.58*10^{-2}$ & $7.0*10^{-3}$ & $8.01\ ***$ \\
  \hline
 11  & $5.69*10^{-2}$ & $6.9*10^{-3}$ & $8.20\ ***$ \\
  \hline
 12  & $4.14*10^{-2}$ & $7.1*10^{-3}$ & $5.86\ ***$ \\
  \hline
 13  & $5.79*10^{-2}$ & $6.9*10^{-3}$ & $8.36\ ***$ \\
  \hline
 14  & $5.17*10^{-2}$ & $7.0*10^{-3}$ & $7.40\ ***$ \\
  \hline
 15  & $4.18*10^{-2}$ & $7.1*10^{-3}$ & $5.93\ ***$ \\
  \hline
 16  & $3.76*10^{-2}$ & $7.1*10^{-3}$ & $5.30\ ***$ \\
  \hline
 17  & $4.86*10^{-2}$ & $7.0*10^{-3}$ & $6.92\ ***$ \\
  \hline
 18  & $5.11*10^{-2}$ & $7.0*10^{-3}$ & $7.31\ ***$ \\
  \hline
 19  & $3.52*10^{-2}$ & $7.1*10^{-3}$ & $4.95\ ***$ \\
  \hline
 20  & $2.96*10^{-2}$ & $7.2*10^{-3}$ & $4.14\ ***$ \\
  \hline
 21  & $3.92*10^{-2}$ & $7.1*10^{-3}$ & $5.54\ ***$ \\
  \hline
 22  & $2.51*10^{-2}$ & $7.2*10^{-3}$ & $3.49\ ***$ \\
  \hline
 23  & $2.70*10^{-2}$ & $7.2*10^{-3}$ & $3.76\ ***$ \\
  \hline
 24  & $3.50*10^{-2}$ & $7.1*10^{-3}$ & $4.93\ ***$ \\
  \hline
 25  & $2.32*10^{-2}$ & $7.2*10^{-3}$ & $3.23\ **$ \\
  \hline
\end{tabular}
\caption{Estimated parameters $\beta_{1i}$ for MSFT asset in the high activity regime. Stars indicate significance levels: $*** \left( 0.001 \right),\ ** \left( 0.01 \right),\ * \left( 0.05 \right),\ . \left( 0.1 \right),\quad  \left( 1 \right) $.}
\end{table} 


\subsection{Comparison with real data}

After having estimated the three models on the real data, we have generated  for each model $25$ data samples of length $10^{6}$ observations. In this way we are be able to determine an empirical statistical error on quantities that we measure
on these artificial samples. We have considered three quantities to be compared with real data. Beside the autocorrelation of squared returns, in order to analyze the return distribution at different transaction time scales $\Delta t$, we have measured the empirical standard deviation and excess kurtosis 
\begin{eqnarray}
\sigma\left(\Delta t\right)=\left(E\left[\left( (p_m(t+\Delta t)-p_m(t))-E[p_m(t+\Delta t)-p_m(t)]\right)^2\right]\right)^{1/2} \\
\kappa\left(\Delta t\right)=\frac{E\left[\left( (p_m(t+\Delta t)-p_m(t))-E[p_m(t+\Delta t)-p_m(t)]\right)^4 \right]}{\sigma^4\left(\Delta t\right)}-3
\end{eqnarray}
The normalized standard deviation $\sigma_{N}\left(\Delta t\right)=\sigma\left(\Delta t\right)/\sqrt{\Delta t}$ gives information of the diffusive character of the price process, because  $\sigma_{N}\left(\Delta t\right)$ is constant for diffusion. The behavior of $\kappa\left(\Delta t\right)$ as a function of $\Delta t$ describes the convergence of the distribution of returns toward the Gaussian distribution \citep{Bouchaud}.   

\begin{figure}
	\centering
		\includegraphics[width=0.49\textwidth]{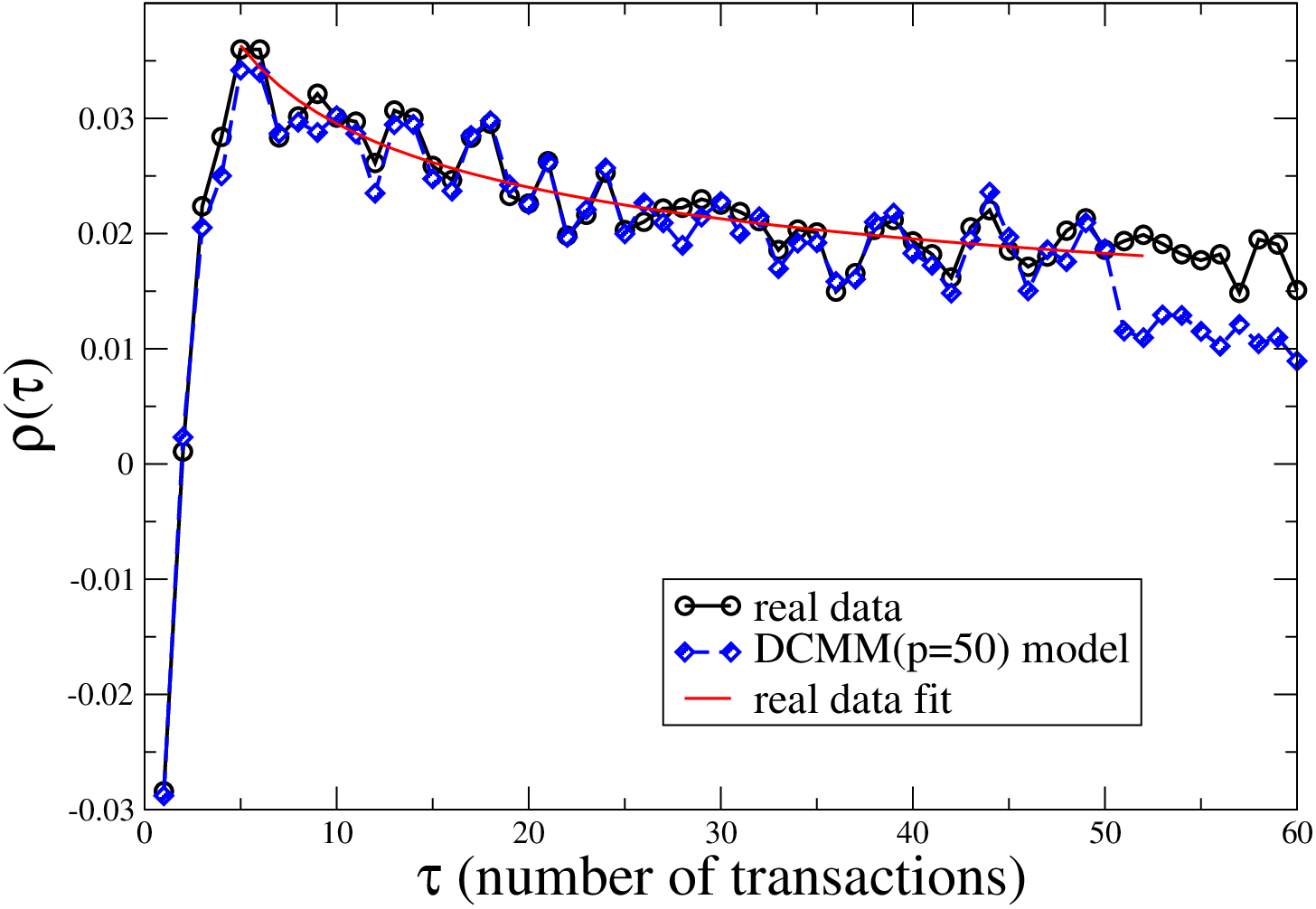}
		\includegraphics[width=0.49\textwidth]{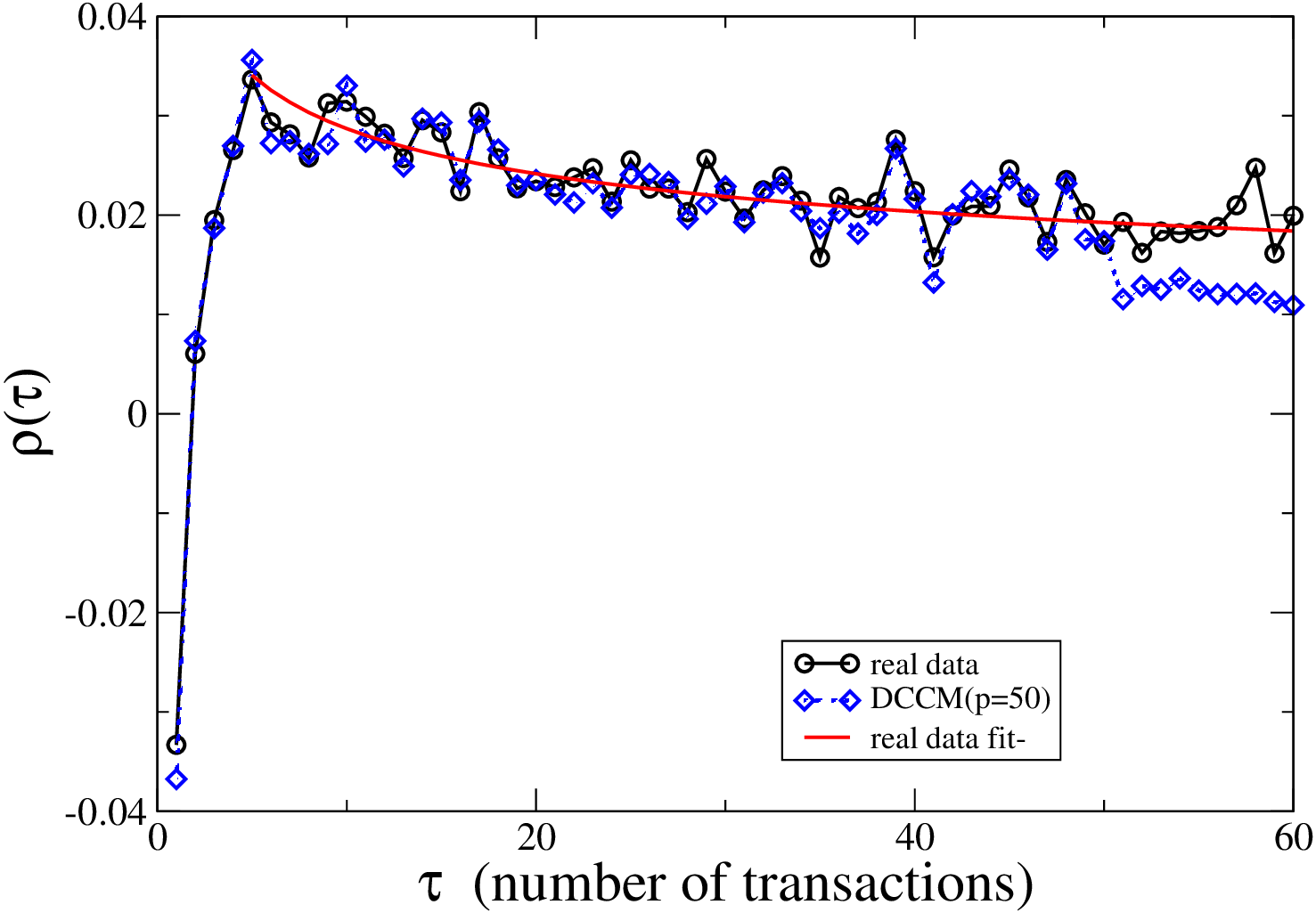}
	\caption{Empirical autocorrelation functions $corr\left(r^{2}\left(t\right),r^{2}\left(t+\tau \right)\right)$ for real (black) and simulated (red) data according to DCMM(50) model. The red squares are a power law fit on the real data. The left panel refers to MSFT and the right panel to CSCO.}
	\label{fig:ACFr2simnew}
\end{figure}

We first investigate the autocorrelation properties of squared returns $\rho\left(\tau \right)$. This function is compatible with zero for MS$_{B}$ and MS models except for the first lag where we have measured a significant positive value $\rho\left(\tau =1\right)\approx0.01$.  
The model with regressors DCMM($p=50$), instead, is able to reproduce remarkably well the values of $\rho\left(\tau \right)$ up to $\tau =50$, as we can see from Fig. \ref{fig:ACFr2simnew}, both for MSFT and for CSCO. The behavior of $\rho\left(\tau \right)$ around $\tau\simeq 0$ is also very well reproduced by the model. 
The model underestimates the  values of the autocorrelation of the real process for $\tau >50$ but it generates values that are still significantly positive. We have performed a 
power law fit on real and DCMM($p=50$) simulated data for values of lags corresponding to $\tau \in \left[6,50\right]$. For real data we found $\alpha=0.298(0.023)$ and for simulated data $\alpha=0.300(0.028)$. Since $\alpha<1$ this model is able to reproduce long memory shape of correlation $\rho\left(\tau \right)$
for a number of values of lags $\tau$ equal to the order of model $p$.

\begin{figure}
\centering
\includegraphics[width=0.45\textwidth]{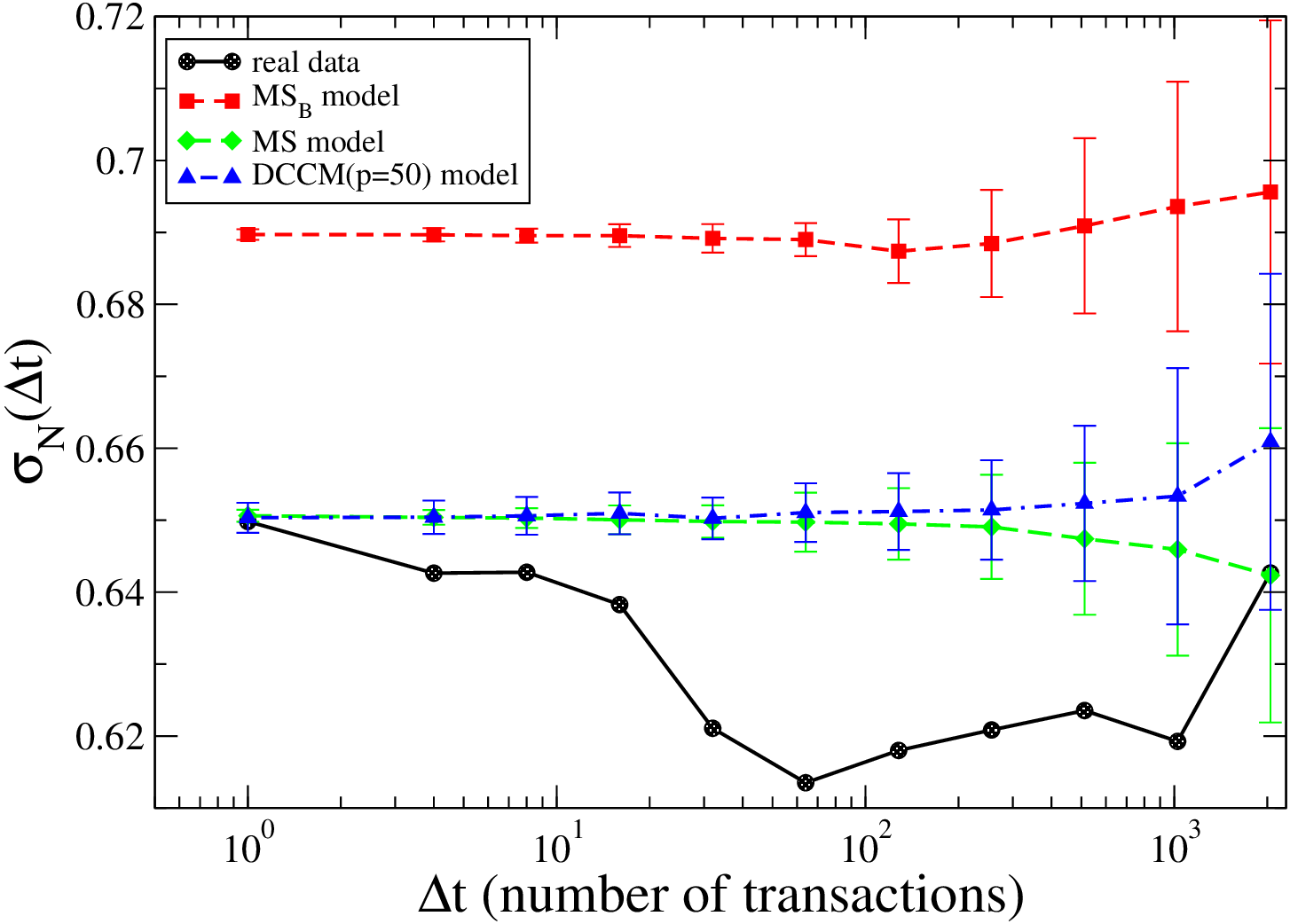}
\includegraphics[width=0.45\textwidth]{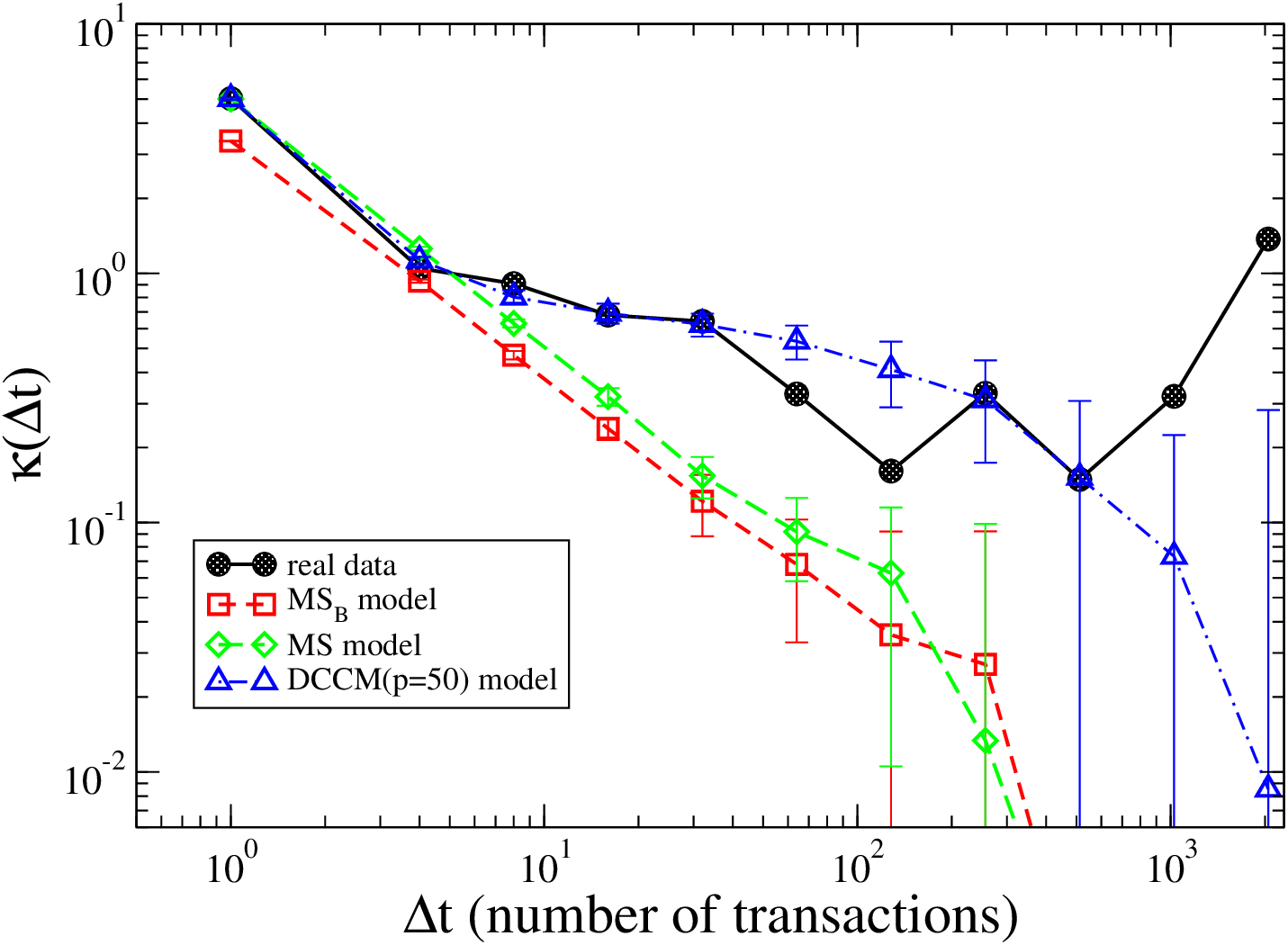}
\caption{Left. Rescaled volatility $\sigma_{N}\left(\Delta t\right)$ of aggregated returns on time scale $\Delta t$ for $MS_{B}$ (red line), $MS$ (green line), and $DCMM(p=50)$ (blue line), compared with the same quantity  for real data for high volatility series (black line).
 Right. Excess kurtosis $\kappa\left(\Delta t\right)$ of aggregated returns on time scale $\Delta t$  for $MS_{B}$ (red line), $MS$(green line), $DCMM(p=50)$ (blue line), compared with the same quantity  for real data for high volatility series (black line). In both panels error bars are the standard deviation obtained from 25 Monte Carlo simulations of the corresponding models. }
\label{fig:KUMSFTASERIES}
\end{figure}

We then analyzed the distributional properties, i.e. normalized standard deviation $\sigma_{N}\left(\Delta t\right)$ and excess kurtosis $\kappa\left(\Delta t\right)$. 
For each value of $\Delta t$ and for each model we calculate the average and standard deviation of the $25$ simulations and we compare the simulation results with real data (see Fig. \ref{fig:KUMSFTASERIES}).

The three models are clearly diffusive. Moreover MS and DCMM($p=50$) models reproduce  the empirical values of $\sigma_{N}$  better than the MS$_{B}$ model.
The difference between MS and DCMM($p=50$) models are appreciable only for $\Delta t >128$, i.e. this parameter is almost the same for these two models. 

The behavior of excess kurtosis, instead, is different between the models (see the right panel of Fig. \ref{fig:KUMSFTASERIES}). The excess kurtosis for MS$_{B}$ and MS models is well fit by a power law $\kappa\left(\Delta t\right) \sim \Delta^{-\alpha}$ with $\alpha=0.901(0.027)$ (MS$_{B}$) and $\alpha=0.997(0.052)$ (MS). These values are consistent with a short range correlation of volatility. In fact, it can be shown \citep{Bouchaud} that stochastic volatility models with short range autocorrelated volatility are characterized by $\alpha=1$. On the contrary, stochastic volatility models with long range autocorrelated volatility display a slower decay. This is exactly what it is observed for real data and for the DCMM($p=50$) model. In both cases we observe an anomalous scaling of kurtosis that is more compatible with a stochastic volatility model in which volatility is a long memory process.

\section{Conclusion}
\label{sec:conclusion}

We have developed Markov-switching models for describing the coupled dynamics of spread and returns of large tick assets in transaction time. The underlying Markov process is the  process of transitions between consecutive spread values. In this way returns are described by different processes depending on whether the spread is constant or not in time. We have shown that this mechanism is needed in order to model the different shape of the distribution of mid-price changes at different aggregation in number of trades. In order to be able to model the persistent volatility clustering, we have introduced a Markov model with logit regressors represented by past values of returns and squared returns.

We have calibrated the model on the stock Microsoft and Cisco and, by using Monte Carlo simulations, we have found that the model reproduces remarkably well and in a quantitative way the empirical stylized facts. In particular we are able to reproduce the shape of the distribution at different aggregations, uncorrelated returns, diffusivity, slowly decaying autocorrelation function of squared returns, and anomalous decay of kurtosis on different time scales, i.e. the convergence to the Gaussian.

As a possible extension, we observe that, if we want to reproduce more precisely the autocorrelation function of squared returns up to a certain number of lags, we need to estimate a number of parameters, i.e. order of model, at least equal to this value. We find that these parameters scale with a power law function of parameter's index, i.e it is a function of the number of past lags at which regressors are defined. A possible improvement of this model could be to develop a model in which we estimate directly a parametric function with a small number of parameters (for example a power law function) that can describe how these parameters scale when we consider a certain order for the model.      

Finally we note that we have developed this model in the case of large tick assets but this limitation is represented only by the choice of a limited set of values for spread and returns variables. In principle the extension to any kind of asset is represented only by a model in which we can have several values for spread, not only $1$ or $2$, and a broader set of values for returns. 

\section*{Acknowledgements}
We would like to thank Alessandro Profeti and Andrea Carlo Giuseppe Mennucci for the development and support of the computer facility HAF922.sns used to perform data analysis and Montecarlo simulations, written in R language, reported in this article.  Authors acknowledge partial support by the grant SNS11LILLB “Price formation, agents het- erogeneity, and market efficiency”

\appendix
\section{Correlation of squared returns for DCMM(p) model}
\label{sec:Appendix A}
The definition of the process for $r^2\left(t\right)\in \left \{ 0,1,4 \right \}$ in the case of a general value of $p$ for the DCMM model is reported in Eq. \ref{eq:1.11}. This stochastic process
is a stationary Markov process of order $p$ for each value of $k$ \citep{Zucchini} :  
\begin{equation}
\label{eq:3.1}
\begin{split}
 P& \left(r^2\left(t\right)=\left(3-i_{p+1}\right)^2|x\left(t\right)=k;\, r^2\left(t-1\right)=\left(3-i_{p}\right)^2 \right. , \\
  & \left. \cdots, r^2\left(t-p\right)=\left(3-i_{1}\right)^2;\boldsymbol{\theta}_{k} \right)=V_{x\left(t\right);i_{1}i_{2}...i_{p+1}}, 
\end{split}
\end{equation}
where we have $k\in \left \{ 1,2,3,4 \right \}$ and a $p+1$-dimensional vector of indices $\hat{\boldsymbol{i}}=\left(i_{1},i_{2},\cdots,i_{p+1}\right)$, 
where each index can assume values $i_{l}\in \left \{ 1,2,3 \right \}$ for each $l\in\left \{1,2,\cdots,p+1\right \}$. 
We stress the concept that the index $i_{p+1}$ defines the present value of the squared return $r^{2}\left(t\right)$, instead the indices $i_{1},\cdots,i_{p}$
define the past history of the process of squared returns, i.e. $i_{1}$ defines the oldest value of $r^{2}=r^{2}\left(t-p\right)$.
The transition probabilities are given by:
\begin{eqnarray}
\label{eq:3.2}
\ V_{x\left(t\right)=k\in\left \{ 1,4 \right \};i_{1}i_{2}...,i_{p+1}=1}&=&\eta_{k}\left(i_{1},\cdots,i_{p}\right)= \frac{exp\left[\alpha_{k}+\sum_{l=1}^p \beta_{k,l}\left(3-i_{p-l+1}\right)^2\right]}{1+exp\left[\alpha_{k}+\sum_{l=1}^p \beta_{k,l}\left(3-i_{p-l+1}\right)^2\right]}, \nonumber \\
\ V_{x\left(t\right)=k\in\left \{ 1,4 \right \};i_{1}i_{2}...,i_{p+1}=2}&=& 0, \nonumber \\
\ V_{x\left(t\right)=k\in\left \{ 1,4 \right \};i_{1}i_{2}...,i_{p+1}=3}&=& \frac{1}{1+exp\left[\alpha_{k}+\sum_{l=1}^p \beta_{k,l}\left(3-i_{p-l+1}\right)^2\right]}, \nonumber \\
\ V_{x\left(t\right)=k\in\left \{ 2,3 \right \};i_{1}i_{2}...,i_{p+1}=1}&=& 0, \nonumber \\
\ V_{x\left(t\right)=k\in\left \{ 2,3 \right \};i_{1}i_{2}...,i_{p+1}=2}&=& 1, \nonumber \\
\ V_{x\left(t\right)=k\in\left \{ 2,3 \right \};i_{1}i_{2}...,i_{p+1}=3}&=& 0,
\end{eqnarray}
for each value of the $p$-dimensional vector $\boldsymbol{i}=\left(i_{1},\cdots,i_{p}\right)$.
We have $3^{p+1}$ values for the transition probabilities whit normalization:
\begin{equation}
\label{eq:3.3}
\forall k;\,\, \forall\ i_{1},\cdots,i_{p}\,\,: \sum_{i_{p+1}=1}^3 V_{x\left(t\right)=k;i_{1}i_{2}...,i_{p+1}}=1. 
\end{equation} 
We can recover an equivalent Markov(1) process defined on vector-states $\boldsymbol{Y}\left(t\right)$. We define a $p$-dimensional vector of squared returns: 
\begin{equation}
\label{eq:3.4}
\boldsymbol{Y}\left(t\right)\left[\boldsymbol{i}\right]=\left(r^2\left(t-p+1\right)=\left(3-i_{1}\right)^2,\cdots,r^2\left(t\right)=\left(3-i_{p}\right)^2\right),
\end{equation}
In this case the index $i_{p}$ defines the present state of the squared return $r^{2}\left(t\right)$.
The vector-process $\boldsymbol{Y}\left(t\right)$ is a first order Markov chain on the state space $\left \{ 0,1,4 \right \}^{p}$, i.e. $\boldsymbol{Y}\left(t\right)$ can assume $3^{p}$ different values.
We define four transition matrices $U_{x\left(t\right)=k}\in\,M_{3^{p},3^{p}}\left(\mathbb{R}\right)$ in order to represent the equivalent Markov process for each possible value of $x\left(t\right)$.
These matrices describe the transition $\boldsymbol{Y}\left(t\right)\rightarrow\boldsymbol{Y}\left(t+1\right)$, that we could represent also by the transition between vectors of indices: $\left(i_{1},\cdots,i_{p}\right)\rightarrow\left(i_{2},\cdots,i_{p+1}\right)$.  
We have to map the transition probabilities $V_{x\left(t\right)=k;i_{1}i_{2}...i_{p+1}}$ to the elements of matrix $U_{k;m,n}$, where $m,n \in\left \{ 1,\cdots,3^{p} \right \}$. We can obtain this by the simple rule:
\begin{eqnarray}
\label{eq:3.5}
\ \left(i_{1},\cdots,i_{p+1}\right)&\rightarrow& \left(m,n\right), \nonumber \\
\ m\left(i_{1},\cdots,i_{p}\right)&=& \left[\sum_{l=1}^{p-1}3^{p-l}\left(3-i_{l}\right)\right]+4-i_{p}, \nonumber \\
\ n\left(i_{2},\cdots,i_{p+1}\right)&=& \left[\sum_{l=1}^{p-1}3^{p-l}\left(3-i_{l+1}\right)\right]+4-i_{p+1}, \nonumber \\
\ U_{x\left(t\right)=k;m,n}&=&V_{x\left(t\right)=k;i_{1}i_{2}...i_{p+1}}. \nonumber \\
\end{eqnarray}
This rules are unable to fill the entire matrix $U_{k;m,n}$, because when we study the Markov process for $\boldsymbol{Y}\left(t\right)$ we have a lot of forbidden transitions, so the elements of matrix
that aren't captured by the above rules have $0$ values. For the case $p=2$ the shape of $U_{k}$ is:
$$
U_{1}=
\begin{pmatrix}
\left[1-\eta_{1}\left(0,0\right)\right] & 0 & \eta_{1}\left(0,0\right) & 0 & 0 & 0 & 0 & 0 & 0\\
0 & 0 & 0 & \left[1-\eta_{1}\left(0,1\right)\right] & 0 & \eta_{1}\left(0,1\right) & 0 & 0 & 0\\
0 & 0 & 0 & 0 & 0 & 0 & \left[1-\eta_{1}\left(0,4\right)\right] & 0 & \eta_{1}\left(0,4\right)\\
\left[1-\eta_{1}\left(1,0\right)\right] & 0 & \eta_{1}\left(1,0\right) & 0 & 0 & 0 & 0 & 0 & 0\\
0 & 0 & 0 & \left[1-\eta_{1}\left(1,1\right)\right] & 0 & \eta_{1}\left(1,1\right) & 0 & 0 & 0\\
0 & 0 & 0 & 0 & 0 & 0 & \left[1-\eta_{1}\left(1,4\right)\right] & 0 & \eta_{1}\left(1,4\right)\\
\left[1-\eta_{1}\left(4,0\right)\right] & 0 & \eta_{1}\left(4,0\right) & 0 & 0 & 0 & 0 & 0 & 0\\
0 & 0 & 0 & \left[1-\eta_{1}\left(4,1\right)\right] & 0 & \eta_{1}\left(4,1\right) & 0 & 0 & 0\\
0 & 0 & 0 & 0 & 0 & 0 & \left[1-\eta_{1}\left(4,4\right)\right] & 0 & \eta_{1}\left(4,4\right)\\ 
\end{pmatrix},
$$
$$
U_{2}=
\begin{pmatrix}
0 & 1 & 0 & 0 & 0 & 0 & 0 & 0 & 0\\
0 & 0 & 0 & 0 & 1 & 0 & 0 & 0 & 0\\
0 & 0 & 0 & 0 & 0 & 0 & 0 & 1 & 0\\
0 & 1 & 0 & 0 & 0 & 0 & 0 & 0 & 0\\
0 & 0 & 0 & 0 & 1 & 0 & 0 & 0 & 0\\
0 & 0 & 0 & 0 & 0 & 0 & 0 & 1 & 0\\
0 & 1 & 0 & 0 & 0 & 0 & 0 & 0 & 0\\
0 & 0 & 0 & 0 & 1 & 0 & 0 & 0 & 0\\
0 & 0 & 0 & 0 & 0 & 0 & 0 & 1 & 0 
\end{pmatrix},\quad
U_{3}=U_{2},\quad U_{4}=U_{1}\left(\eta_{1}\rightarrow \eta_{4}\right).
$$
In $U_{1}$ we have $\eta_{1}\left(i_{1},i_{2}\right)=\eta_{1}\left(r^2\left(t-2\right)=\left(3-i_{1}\right)^2,r^2\left(t-1\right)=\left(3-i_{2}\right)^2\right)$.
Finally, we define an overall Markov process for $\boldsymbol{Y}\left(t\right)$, defined by 
$4+2p$ parameters: $p_{11}, p_{21}, \alpha_{k}, \boldsymbol{\beta}_{k}'$, where $k \in \left \{ 1,4\right \}$:
\begin{equation}
\label{eq:3.6}
S=\sum_{k=1}^4 \lambda_{k} U_{k},
\end{equation}
where $\lambda_{k}$ are given by Eq. \ref{eq:1.5}.
Now our goal is to calculate the moments for the variable $r^{2}\left(t\right)$ from the process defined by Eq. \ref{eq:3.6}.
First of all we have to solve the eigenvalue equation for $S$ relative to the eigenvalue $1$ in order to determine the stationary probability vector for $\boldsymbol{Y}\left(t\right)$:
\begin{equation}
\label{eq:3.7}
S'\boldsymbol{\Psi}=\boldsymbol{\Psi}.
\end{equation}
The $3p$-dimensional vector $\boldsymbol{\Psi}$ represents all possible values of the stationary $3p$-variate distribution of the variable
$\boldsymbol{Y}\left(t\right)$:
\begin{equation}
\label{eq:3.8}
P\left(\boldsymbol{Y}\left(t\right)\left[i_{1},\cdots,i_{p}\right]\right)=\Psi_{m\left(i_{1},\cdots,i_{p}\right)}.
\end{equation} 
From the $3p$-dimensional vector $\boldsymbol{\Psi}$ we compute the stationary $3$-dimensional probability vector $\boldsymbol{\psi}'=\left(\psi_{1},\psi_{2},\psi_{3}\right)$ for the process $r^{2}\left(t\right)$,
i.e. we have for each index $i_{p}\in \left \{1,2,3\right \}$: 
\begin{equation}
 \label{eq:3.9}
\psi_{i_{p}}=P\left[r^{2}\left(t\right)=\left(3-i_{p}\right)^{2}\right]=\sum_{i_{1}=1}^3 \cdots \sum_{i_{p-1}=1}^3 \Psi_{m\left(i_{1},\cdots,i_{p}\right)},
\end{equation}
where $i_{p}$ defines the present value of $r^{2}\left(t\right)$ and we use mappings defined in Eq. \ref{eq:3.5}. The stationary probability to have a fixed value of $r^{2}$ at time $t$ depends on all possible values
of $r^{2}$ during the past $p-1$ lags. In order to determine the present probabilities we have to sum probabilities corresponding to all possible past trajectories defined
by the past $p-1$ lags.
 
We compute $corr\left(r^{2}\left(t\right),r^{2}\left(t+\tau \right)\right)=\rho\left(\tau\right)$
by means of the transition probabilities  $P\left(r^2\left(t\right)=\left(3-a\right)^2,r^2\left(t+\tau\right)=\left(3-b\right)^2\right)$, 
where $a,b\in \left \{1,2,3\right \}$, of the $p$-order Markov process in term of the matrix $S$:
\begin{equation}
 \label{3.10.1}
 P\left(r^2\left(t\right)=\left(3-a\right)^2,r^2\left(t+\tau\right)=\left(3-b\right)^2\right)=P\left(\boldsymbol{i}\left(a\right),\boldsymbol{j}\left(b\right)\right),
\end{equation}
where $\boldsymbol{i}\left(a\right)=\left(i_{1},\cdots,i_{p}=a\right)$ and $\boldsymbol{i}\left(b\right)=\left(i_{1},\cdots,i_{p}=b\right)$ are the $p$-dimensional
vectors of indices describing the past $p-1$ lags respect to times $t$ and $t+\tau$.
We have to perform the sum of probabilities corresponding to each of the possible values of $i_{1},\cdots,i_{p-1}$ and $j_{1},\cdots,j_{p-1}$, i.e. on $i_{l},j_{l}
\in \left \{1,2,3\right \}$ $\forall l \in \left \{1,\cdots,p-1\right \}$:
\begin{eqnarray}
\label{eq:3.10.2}
\ P\left(\boldsymbol{i}\left(a\right),\boldsymbol{j}\left(b\right)\right)&=&\sum_{\left(i_{1},\cdots,i_{p-1},i_{p}=a\right)} \sum_{\left(j_{1},\cdots,j_{p-1},j_{p}=b\right)} 
  P\left(\boldsymbol{Y}\left(t\right)\left[\boldsymbol{i}\left(a\right)\right],\boldsymbol{Y}\left(t+\tau\right)\left[\boldsymbol{j}\left(b\right)\right]\right) \nonumber \\  
\  &=&\sum_{\left(i_{1},\cdots,i_{p-1},i_{p}=a\right)} \sum_{\left(j_{1},\cdots,j_{p-1},j_{p}=b\right)}  \left(S^{\tau}\right)_{m\left(\boldsymbol{i}\left(a\right)\right),n\left(\boldsymbol{j}\left(b\right)\right)}
   \Psi_{m\left(\boldsymbol{i}\left(a\right)\right)},  
\end{eqnarray}
where we use mappings defined in Eq. \ref{eq:3.5} and the matrix power $S^{\tau}$, because we sum on all possible transitions $\boldsymbol{Y}\left(t\right)\rightarrow \boldsymbol{Y}\left(t+\tau \right)$ holding fixed
the values of indices $i_{p}=a$ and $j_{p}=b$.
At this point we can compute the moments of our interest:
\begin{eqnarray}
\label{eq:3.11}
\ E\left[r^2\left(t\right)\right]&=& \sum_{i=1}^3 \left(3-i\right)^2\psi_{i} =4\psi_{1}+\psi_{2}, \nonumber \\
\ E\left[r^4\left(t\right)\right]&=& \sum_{i=1}^3 \left(3-i\right)^4\psi_{i} =16\psi_{1}+\psi_{2}, \nonumber \\  
\ E\left[r^2\left(t\right)r^2\left(t+\tau\right)\right]&=&\sum_{a=1}^3\sum_{b=1}^3 \left(3-a\right)^2\left(3-b\right)^2 P\left(\boldsymbol{i}\left(a\right),\boldsymbol{j}\left(b\right)\right),  
\end{eqnarray}
from which we can determine the function $\rho\left(\tau\right)$. We have determined the function $\rho\left(\tau\right)$ for $p=3$ making the following
approximation for the matrix $V_{4}$:
\begin{eqnarray}
\label{eq:3.12}
\ V_{x\left(t\right)=k=4;i_{1}i_{2}...i_{p+1}=1}&=& 2\theta_{4}, \nonumber \\
\ V_{x\left(t\right)=k=4;i_{1}i_{2}...i_{p+1}=2}&=& 0, \nonumber \\
\ V_{x\left(t\right)=k=4;i_{1}i_{2}...i_{p+1}=3}&=& 1-2\theta_{4},
\end{eqnarray}
this approximation is justified only in the case $\lambda_{1}\approx 1$, i.e. we have the same approximation that leads us to Eq.  \ref{eq:1.13}.
In this way we have found the results reported in Fig. \ref{fig:DCMMlogit} for DCMM($p=3$).




\begin{thebibliography}{1}
\bibitem{Wallach} Wallach, H.M., 2004. Conditional Random Fields: An Introduction. University of Pennsylvania CIS Technical Report MS-CIS-04-21.
\bibitem{Rabiner} Rabiner, L.R., 1989. A Tutorial on Hidden Markov Models and Selected Applications in Speech Recognition. Proceedings of the IEEE 77, (2), 257-286.
\bibitem{Hamilton 1} Hamilton, J.D., 1994. Time Series Analysis. Princeton University Press, Princeton, New Jersey.
\bibitem{Bouchaud} Bouchaud, J.-P., Potters, M., 2003. Theory of Financial Risks: From Statistical Physics to Risk Management. Cambridge University Press, New York.
\bibitem{McKenzie} McKenzie, E., 2000. Discrete Variates Time Series. University of Strathclyde.
\bibitem{Fokianos} Kedem, B., Fokianos, K., 2002. Regression Models for Time Series Analysis. Wiley-Interscience, Hoboken, New Jersey.
\bibitem{Hautsch} Grob-Klubmann, A., Hautsch, N., 2011. Predicting Bid-Ask
Spreads using long memory autoregressive conditional Poisson Models. Working Paper Humboldt-Universit\"{a}t zu Berlin.
\bibitem{Gillemot} Gillemot, L., Farmer, J.D., Lillo, F., 2006. There's more
to volatility than volume. Quantitative Finance 6 (5), 371-384.
\bibitem{Hamilton} Hamilton, J.D., 2005. Regime-Switching models. Palgrave Dictionary of Economics.
\bibitem{Liesenfeld} Liesenfeld, R., Nolte, I., Pohlmeier, W., 2003.
Modeling financial transaction price movements: a dynamic integer count data
model. Empirical Economics 30, 795-825.
\bibitem{Dayri} Al Dayri, K.A., 2011. Market microstructure and modeling of
the trading flow. These de doctorat, Ecole Polytechnique.
\bibitem{Clauset} Clauset, A., Shalizi, C.R., Newman, M.E.J., 2009. Power-Law
distributions in empirical data. SIAM Review 51 (4), 661-703.
\bibitem{Munnix} Munnix, M.C., Schafer, R., G\"{u}hr, T., 2010. Impact of the
tick-size on financial returns and correlations. Physica A 389 (21), 4828-4843.
\bibitem{Onnela} Onnela, J.-P., Toyli, J., Kaski, K., 2009. Tick size and
stock returns. Physica A 388, 441-454.
\bibitem{LaSpada} La Spada, G., Farmer, J.D. and Lillo, F., 2011. Tick
size and price diffusion, Econophysics of order-driven markets. Springer, 173-187.
\bibitem{LaSpada13} La Spada, G. and Lillo, F., 2013. The effect of round-off error on long memory processes, Studies in Nonlinear Dynamics and Econometrics (in press).
\bibitem{Ferland} Ferland, R., Latour, A. and Oraichi, D., 2004. Integer-valued
GARCH process. Journal of time series analysis 27 (6), 923-942.
\bibitem{Ponzi} Ponzi, A., Lillo, F. and Mantegna, R.N., 2009. Market
reaction to a bid-ask spread change: A power-law relaxation dynamics.
Physical Review E 80 (1), 016112-1/016112-12.
\bibitem{Wyart} Wyart, M., Bouchaud, J.-P., Kockelkoren, J., Potters, M.
, Vettorazzo, M., 2008. Relation between bid-ask spread, impact and
volatility in order-driven markets. Quantitative Finance 8 (1), 41-57.
\bibitem{Rosenbaum} Robert, C.Y. and Rosenbaum, M., 2011. A new approach for
the dynamics of ultra high frequency data: the model with uncertainty zones.
Journal of Financial Econometrics 9, 344-366.
\bibitem{Andersen} Andersen, T.G., Bollerslev, T., 1997. Intraday periodicity and
volatility persistence in financial markets. Journal of empirical
finance 4 (2), 115-158.
\bibitem{Dayri2} Dayri, K., Rosenbaum, M., 2012. Large tick assets: implicit spread and optimal tick size. arXiv:1207.6325.
\bibitem{Eisler} Eisler, Z., Bouchaud, J.P. and Kockelkoren, J., 2012. The price impact of
order book events: market orders, limit orders and cancellations.
Quantitative Finance 12 (9).
\bibitem{Mike} Mike, S., Farmer, J.D., 2008. An empirical behavioral model of liquidity and volatility. Journal of Economic
Dynamics and Control 32, 200-234.
\bibitem{Berchtold1} Berchtold, A., 1999. The double chain Markov model. Communications in Statistics - Theory and Methods 28 (11), 2569-2589.
\bibitem{Berchtold2} Berchtold, A., 2002. High-order extensions of the Double Chain Markov Model. Stochastic Models 18 (2), 193-227.
\bibitem{Raftery} Berchtold, A. and Raftery, A.E., 2002. The mixture transition distribution model for high-order Markov chains and non-gaussian time series. 
Statistical Science 17 (3), 328-356.
\bibitem{Cogburn} Cogburn, R., 1984. The ergodic theory of Markov chains in random environments. Zeitschrift f\"{u}r Wahrscheinlichkeitstheorie und Verwandte Gebiete 66 (1), 109-128.
\bibitem{Cogburn2} Cogburn, R., 1990. On direct convergence and periodicity for transition probabilities of Markov chains in random environments. The Annals of probability 18 (2), 642-654.   
\bibitem{Schnatter} Fr\"{u}hwirth-Schnatter, S., 2006. Finite mixture and Markov switching models. Springer series in statistics, .
\bibitem{Zucchini} Zucchini, W. and MacDonald, I.L., 2009. Hidden Markov Models for Time Series: an Introduction Using R. Chapman \& Hall/CRC, Taylor \& Francis Group.
\bibitem{Timmermann} Timmermann, A., 2000. Moments of Markov switching models. Journal of Econometrics 96, 75-111.
\bibitem{Ryden} Ryden, T., Terasvirta, T., Asbrink, S., 1998. Stylized facts of daily returns series and the hidden Markov model. 
Journal of Applied Econometrics 13 (3), 217-244.
\bibitem{Bulla} Bulla, J., Bulla, I., 2006. Stylized facts of financial time series and hidden semi-Markov models. 
Computational statistics \& data analysis 51, 2192-2209.
\bibitem{Fitzpatrick} Fitzpatrick, M., Marchev, D., 2012. Efficient Bayesian estimation of the multivariate Double Chain Markov Model. 
Statistics and Computing, Springer.
\bibitem{Eisenkopf} Eisenkopf, A., 2008. The real nature of credit transitions. Working paper, URL: http://ssrn.com/abstract=968311.
\bibitem{Granger} Granger, C.W.J., 1972. Infinite variance and research strategy in time series analysis.
Journal of American Statistical Association 67, 275-285.
\bibitem{Clark} Clark, P.C., 1973. A subordinated stochastic process model with finite variance for speculative prices. 
Econometrica 41, 135-155.
\bibitem{Engle} Engel, C., Hamilton J.D., 1990. Long swings in the dollar: are they in the data and do markets know it?
American Economic Review 89, 689-713.
\bibitem{Hamilton2} Hamilton, J.D., 1989. A new approach to the economic analysis of nonstationary time series and the business cycle.
Econometrica 57, 357-384.
\bibitem{Pham} Guilbaud, F., Pham H., 2011. Optimal high frequency trading with limit and market orders. arXiv:1106.5040v1.
\bibitem{Plerou}  Plerou, V., Gopikrishnan, P. and  Stanley, H. E., 2005. 
Quantifying fluctuations in market liquidity: Analysis of the bid-ask spread. Physical Review E Vol. 71, 046131.
\bibitem{Bouchaudnew} Gareche, A., Disdier, G., Kockelkoren, J., and  Bouchaud, J.-P. 2013. A Fokker-Planck description for the queue dynamics of large tick stocks. Preprint at arXiv:1304.6819.


\end{thebibliography}
\end{document}